\documentclass{aa}
\usepackage{graphicx}
\usepackage{psfig}

\begin{document}

\title{SINFONI adaptive optics integral field spectroscopy of
  the Circinus Galaxy
\thanks{Based on observations collected at the European Southern
        Observatory, Chile}
}
\author{}
\author{F. Mueller S\'anchez \inst{1} \and R.~I. Davies \inst{1} 
  \and F. Eisenhauer \inst{1} \and L.~J. Tacconi \inst{1}
  \and R. Genzel \inst{1, 2} \and A. Sternberg \inst{3}
  }
\institute{}
\institute{Max-Planck-Institut f\"ur extraterrestrische Physik,
        Giessenbachstrasse, Postfach 1312, D-85741 Garching, Germany
        \and Department of Physics, 366 Le~Conte Hall, University of 
        California, Berkeley, CA, 94720-7300, United States
	\and School of Physics and Astronomy, Tel Aviv University, Tel
	Aviv 69978, Israel
}

\offprints{F. Mueller S\'anchez \\
    \email{frankmueller@mpe.mpg.de}}

\date{Received / Accepted}

\abstract
{}
{To investigate the star formation activity and the gas and stellar dynamics 
on scales of a few parsecs in the nucleus of the Circinus Galaxy.}
{Using the adaptive optics near infrared integral field 
spectrometer SINFONI on the VLT, we have obtained observations of the 
Circinus galaxy on scales of a few parsecs and at a spectral
resolution of 70\,km\,s$^{-1}$ FWHM. The physical properties of the nucleus 
are analyzed by means of line and velocity maps extracted from the 
SINFONI datacube. 
Starburst models are constrained using the Br$\gamma$ flux, stellar
continuum (as traced via the CO absorption bandheads longward of
2.3$\mu$m), and radio continuum.}
{The similarity of the morphologies of the H$_2$ 1-0\,S(1) 2.12$\mu$m and 
Br$\gamma$ 2.17$\mu$m lines to the stellar continuum and also their
kinematics, suggest a common origin in star formation.
Within 8\,pc of the AGN we find there has been a recent starburst
in the last 100\,Myr, which currently accounts for 1.4\% of the
galaxy's bolometric luminosity.
The similarity of the spatial scales over which the stars and gas exist
indicates that this star formation is occuring within the torus;
and comparison of the gas column density through the torus to the maximum
possible optical depth to the stars implies the torus is a clumpy
medium.
The coronal lines show asymmetric profiles with a spatially compact
narrow component and a spatially extended blue wing.
These characteristics are consistent with some of the emission arising
in clouds gravitationally bound to the AGN, and some outflowing in
cloudlets which have been eroded away from the bound clouds.}
{}

\keywords{galaxies: active -- 
 galaxies: individual: Circinus -- 
 galaxies: nuclei -- 
 galaxies: Seyfert -- 
 galaxies: starburst -- 
 infrared: galaxies}

\titlerunning{SINFONI Spectroscopy of the Circinus galaxy}

\maketitle

%------------------------------------------------------------------------------

\section{Introduction}

In the context of active galactic nuclei, star formation activity and
the gas and stellar dynamics on scales of a few parsecs to a few tens
of parsecs can be counted among the main debated issues.
The unified model of active galaxies (see Lawrence \cite{lawrence} for
a review) 
assumes that the inner region of Seyfert 2 galaxies is comprised of a
dense circumnuclear  
torus that hides the nucleus and the broad-line region from our line of sight 
at near-infrared and optical wavelengths. The size scales on which
models predict  
the canonical torus vary from an inner edge at 1 pc out to several
tens of parsecs  
(Pier \& Krolik \cite{pierandkrolik}, Nenkova et al. \cite{nenkova},
Schartmann et al. \cite{sch05}). 
These crucial size scales are exactly those that can be resolved 
with SINFONI in the nearest AGN. 

The Circinus galaxy, at a distance of $4.2\pm0.8$ Mpc (Freeman et
al. \cite{freeman}), 
is an ideal subject to study because of its proximity ($1\arcsec = 20$ pc). 
It is a large, highly inclined ($i=65 \degr$ Freeman et
al. \cite{freeman}), spiral galaxy that hosts both 
a typical Seyfert 2 nucleus and a circumnuclear starburst on scales of 100--200\,pc 
(Maiolino et al. \cite{maio98}). 
Evidence for an obscured Seyfert 1 nucleus is provided by the finding
of a broad  
($\mathrm{FWHM} = 3300\, \mathrm{km s^{-1}}$) $\mathrm{H\alpha}$ line
component in  
polarized light (Oliva et al. \cite{oliva98}).
Their picture is also supported by recent X-ray observations. 
Those above 10 keV suggest direct X-ray detection of the nucleus through a column density of 
$4\times10^{24}\, \mathrm{cm^{-2}}$ (Matt et al. \cite{matt99}). 
The X-ray spectrum below 10 keV exhibits a flat continuum and a very prominent iron line, 
indicative of Compton scattering and fluorescent emission from gas 
illuminated by an obscured X-ray continuum source (Matt et al. \cite{matt96}). 
Circinus shows highly ionized gas extending along the minor axis 
of the galaxy, with a morphology that is reminiscent of the ionization cones seen 
in other Seyfert galaxies (Marconi et al. \cite{marconi94a}). 
H$_2$O maser emission has been detected, and the masing gas traces a thin 
accretion disk about 0.4 pc in radius, with, in addition, a fraction 
of the masers originating outside the disk, in what appears 
to be an outflow within $\sim$1 pc of the nucleus and aligned with 
the ionization cone (Greenhill et al. \cite{greenhill03}). 
Optical and near-infrared spectrophotometry of the nucleus show 
a typical Seyfert spectrum, including strong coronal lines 
(Oliva et al. \cite{oliva94}; Prieto et al. \cite{pri04}). 
There are also lines from H$_2$ and low-excitation ionic species, 
both believed to be associated with starforming regions. 
The distribution and kinematics of the Br$\gamma$ line have been 
interpreted in terms of ongoing star formation activity within a few 
tens of parsecs of the active nucleus (Maiolino et al. \cite{maio98}). 
A young stellar population with an age between $4\times10^7$ and $1.5\times10^8$
was found between these scales. Recent observations of the Circinus Galaxy in the 
range between $1-10\, \mu$m (Prieto et al. \cite{pri04}), 
resolve a $Ks$-band source with a FWHM of 
$\sim2$ pc and a spectral energy distribution 
compatible with a dust temperature of 300 K.   

This paper presents high-resolution, nearly diffraction-limited
integral field spectroscopic data of the Circinus galaxy in the
$K$-band observed with the adaptive optics (AO) assisted
imaging spectrograph SINFONI (\cite{bon04}, 
\cite{eis03b}). The physical scale 
associated with our spatial resolution is 4\,pc, allowing us to
investigate the properties of the nucleus  
and its interaction with the circumnuclear environment, highlighting
the relationship between star formation and galactic nuclear
activity.

We describe our observations and the data reduction procedure 
in Sect.~\ref{observations}. Sect.~\ref{results} describes the features 
observed in the spectrum of the galaxy and the methods we used to estimate 
the spatial resolution of our observations.
In Sect.~\ref{dustemission} 
we analyze the morphology of the nuclear source and the origin of the observed 
$K$-band continuum. Sect.~\ref{starformation} discusses the morphologies 
of the emission and the absorption, and the age of the starburst at
the nucleus.  
Then, in Sect.~\ref{kinematics} we discuss the kinematics 
of the emission and the mass distribution in the galaxy. 
Following these results, 
Sect.~\ref{density} deals with the gas density at the nucleus and the
physical characteristics of the torus. 
A study of the detected coronal lines is presented 
in Sect.~\ref{coronal}, and we present our conclusions in Sect.~\ref{concl}.

\section{Observations and data reduction} \label{observations}

The data presented here were obtained on 15~Jul~2004 during
commissioning of SINFONI (\cite{bon04, eis03b}) on the VLT UT4.
The instrument consists of a cryogenic near infrared integral field
spectrometer SPIFFI (\cite{eis03a}) coupled to a visible curvature
adaptive optics (AO) system (\cite{bon03}).
SINFONI performs imaging spectroscopy by cutting the two-dimensional 
field of view into 32 slices and then rearranging each of the slices 
(slitlets) onto a one-dimensional pseudo longslit, which is dispersed by
a grating wheel and at last the spectra are imaged on the detector.

The AO module was able to correct on the nucleus of Circinus (for
which it measured R=14.1\,mag) in
seeing of $\sim$0.5\arcsec, to reach a resolution of 0.2\arcsec\ in
the $K$-band (see Section~\ref{dustemission}).
This performance is good considering that the optical
nucleus of Circinus is rather extended rather than point-like.
With the appropriate pixel scale selected, the spectrograph was able,
in a single shot, to obtain spectra covering the whole of the $K$-band
(approximately 1.95--2.45 $\mathrm{\mu m}$) at a spectral resolution of
$R\sim4200$ for each $0.0125\arcsec\times0.025\arcsec$ pixel in a
$0.80\arcsec\times0.80\arcsec$ field of view.
A total of 6 sky and 6 on-source exposures of 300\,sec each were
combined to make the final data cube with a total integration time of 1800\,sec.

The data were reduced using the SINFONI custom reduction package
SPRED (Abuter et al. \cite{abuter05}).
This performs all the usual steps needed to reduce near infrared
spectra, but with the additional routines for reconstructing the data
cube.
Following subtraction of the sky frames from the on-source frames, the
data were flatfielded and corrected for dead/hot pixels.
The data were then interpolated to linear wavelength and spatial
scales, after which the slitlets were aligned and stacked up to
create a cube.
Finally the atmospheric absorption was compensated using the A0V star
HD\,190285.
Flux calibration was performed by comparison to the high spatial
resolution broad band K$_s$ image and photometry obtained with
NAOS-CONICA of 11.4 mag in a 0.38\arcsec aperture (Prieto et al. \cite{pri04}), 
which was also cross-checked with 2MASS photometry.

No additional point-spread function (PSF) calibration frames using
stars were taken.
This is primarily because, although in principle one can match the
brightness of a calibration star on the wavefront sensor to the AGN, it
is not possible to replicate either the spatial extent of the AGN or
the background galaxy light associated with it -- resulting in a
potentially considerable mismatch between the science and calibration
PSFs (\cite{dav04}).
The spatial resolution has instead been measured using the methods
described in Section~\ref{psf}.

\section{{Spectroscopic features and spatial resolution}} \label{results}

\begin{table}
        \begin{center}
        \begin{tabular}{l c c}
        \hline 
        \hline \noalign{\smallskip}
        Line &   $\lambda^a$      & Flux$^b$ \\
             & ($\mathrm{\mu m}$) & ($10^{-18}$ $\mathrm{W m^{-2}}$)\\
        \hline \noalign{\smallskip} 
        $\mathrm{H_2}$ 1-0\,S(3) & 1.9576 &  6.0 \\
        $\mathrm{[Si}${\sc vi}] & 1.9634 & 53.9  \\
        $\mathrm{H_2}$ 1-0\,S(2) & 2.0338 &  3.8  \\
        $\mathrm{[Al}${\sc ix}] & 2.040  & 10.0  \\
        $\mathrm{[He}${\sc i}]  & 2.0587 &  5.0  \\
        $\mathrm{H_2}$ 1-0\,S(1) & 2.1218 &  9.0  \\
        $\mathrm{Br\gamma}$     & 2.1661 & 14.1  \\
        $\mathrm{H_2}$ 1-0\,S(0) & 2.2233 &  3.4  \\
        $\mathrm{[Ca}${\sc viii}]&2.3213 & 36.9  \\
        $\mathrm{H_2}$ 1-0\,Q(1) & 2.4066 &  3.4  \\
        \hline
        \hline
        \end{tabular}
        \end{center}

$^a$ Wavelengths are in the rest frame \\
$^b$ Uncertainities are approximately $3.2\times10^{-20}$ $\mathrm{W m^{-2}}$ \\

        \caption{Measured nuclear emission line fluxes for Circinus.
                 All measurements are given for a 0.8\arcsec\ circular aperture
                 centered on the continuum peak.}
\label{fluxes}
\end{table}

The purpose of this section is to present general results extracted from the data to provide 
an overview of the features observed in the nuclear region of Circinus. 
A more detailed analysis is given in subsequent sections.

\subsection{Spatial Resolution}  \label{psf}

A good way to estimate the spatial resolution is by using a broad line 
in the spectrum, such as a potential observation of Br$\gamma$
emission from the BLR, which at the distance of Circinus is expected
to be spatially unresolved. 
This has the advantage of measuring the spatial resolution directly from 
the science frames, and includes all effects associated with shifting 
and combining the cube.
Nevertheless, as can be seen in Figure~\ref{spectrumcircinus}, we have
detected only narrow lines, which are  
spatially resolved in our observations. 
As a result, and due to the lack of other point
sources in the field, we used indirect means to derive our spatial
resolution, making use of a NACO image of the galaxy from 
Prieto et al. (\cite{pri04})
which contains unresolved star clusters over a field of view of 
$27\arcsec\times27\arcsec$.
%The adopted pixel scale of this image is 0.027 arcsec/pixel. 
Two approaches were used to estimate the PSF of our observations 
by means of the NACO image.
One method of estimating the SINFONI PSF effectively consists of
deconvolving the SINFONI continuum image with the intrinsic image of
the galaxy, since:
\[
%\begin{equation}
        \mathrm{IM_S} = 
                \mathrm{PSF_S} \otimes 
                        \mathrm{IM_{intr}}
%\end{equation}
\]
where IM$\mathrm{_S}$ is the Circinus image from SINFONI, PSF$\mathrm{_S}$
is the point spread function, and IM$\mathrm{_{intr}}$ is the intrinsic image
of the galaxy. We obtained the intrinsic image in a similar way, 
by first deconvolving the NACO image with the NACO PSF. 
In the NACO image, the smooth underlying galaxy profile was
approximated by isophotal analysis using  the isophote package from IRAF. 
This was then subtracted, leaving only the compact sources in the field.
The narrowest point in this subtracted image was used as the NACO PSF, 
which was measured directly to have 0.14\arcsec\ FWHM. 
The deconvolution of the NACO image with its PSF
was performed using the Lucy algorithm (Lucy \cite{lucy74})
also implemented in IRAF.
This deconvolved image was rebinned to our pixel scale of 0.0125\arcsec/pix.
For the final stage, instead of deconvolving IM$\mathrm{_S}$ with 
IM$\mathrm{_{intr}}$, we fit the SINFONI image with
the intrinsic image convolved with the parameterized PSF$\mathrm{_S}$.
This avoids the noise amplification which is an inherent feature of
deconvolution. The PSF$\mathrm{_S}$ could be well matched by
a symmetrical moffat function.
Mismatches were minimized by varying the parameters (center, scale,
$\alpha$ and $\beta$) of a moffat function defined as 
$\mathrm{I}(r) = \big(1+(\frac{r}{\alpha})^2\big)^{-\beta}$ 
and then performing a 
$\chi^2$ minimization test, resulting in a FWHM spatial resolution of
$0.20\arcsec$ ($\sim 3.9$ pc) with $\alpha=0.148\arcsec$ and $\beta=1.8$ for this method.  
An elliptical moffat function was also tested as a possible PSF for
our observations.
However, this did not improve significantly the fit and so was discarded.

In order to avoid the first deconvolution of the method described above, 
the second approach to derive the SINFONI PSF consists of convolving the 
NACO PSF with a parameterized degradation function, the latter being obtained by  
fitting the SINFONI image with
the NACO image convolved with the same function. A symmetrical 
moffat function was used as degradation function. 
Once again, mismatches were minimized by varying the parameters 
of the moffat function and then performing a 
$\chi^2$ minimization.
The SINFONI PSF was obtained by convolving the NACO PSF with the
degradation function. 
The final FWHM resolution estimated in this 
way was $0.22\arcsec$ ($\sim 4.2$ pc) with $\alpha=0.219\arcsec$ and $\beta=3.2$.

The confidence on the resulting PSF for both cases was evaluated by the residual of 
the fitting the following function which considers the difference
between the data and model, weighted according the flux in each pixel,
and normalised by the total flux in the frame:
\[
%\begin{equation} 
\textrm{Residual}=\frac{\sqrt{\sum((\textrm{IM$_{\rm S}$}-\textrm{Model})^2 \cdot 
     \textrm{IM$_{\rm S}$})}}{\left(\sum(\textrm{IM$_{\rm S}$})\right)^{1.5}} \cdot 100
%\end{equation} 
\]
where $IM_{\rm S}$ is the SINFONI continuum image as above and $Model$
is our constructed image -- either the intrinsic galaxy profile
convolved with the SINFONI PSF, or the NACO image convolved with the
degradation function.
The numerator is the quantity minimized during 
the fit, which was then normalized using the integrated image of the
continuum.
The residual of the fit in both cases was evaluated as $0.1\%$, indicating 
that both methods performed equally well in estimating the PSF. 
We chose to adopt the PSF obtained by the second approach  
because no deconvolution is performed.

\subsection{Distribution of emission and absorption features} \label{spectrum}

\begin{figure*}
    \begin{center}
     \resizebox{0.8\textwidth}{!}{\includegraphics[clip]{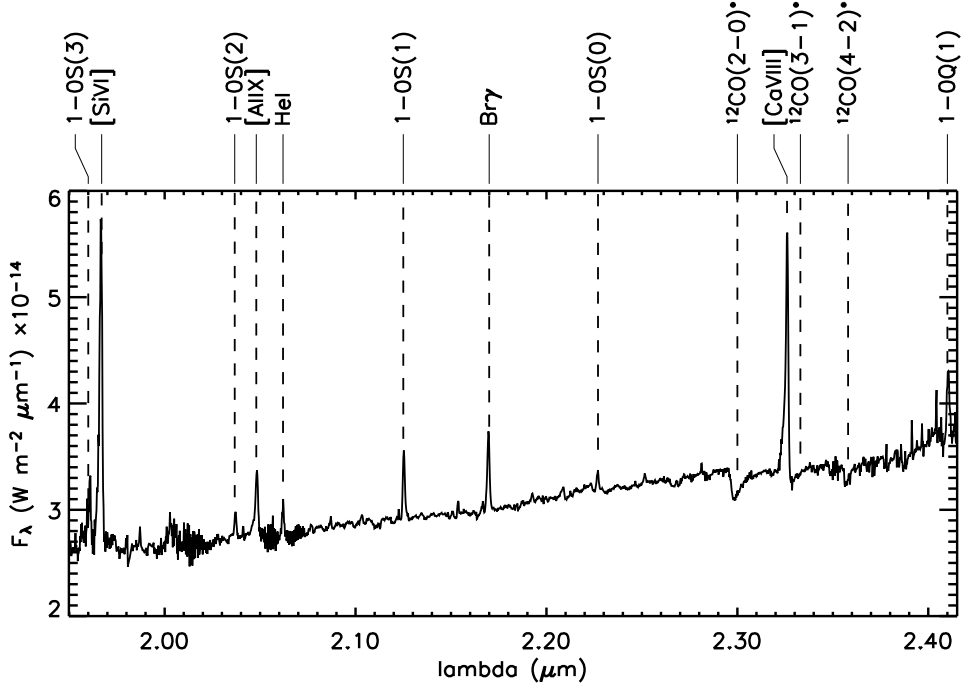}}   
    \caption{Nuclear spectrum of Circinus, extracted from the SINFONI datacube
             in a 0.8\arcsec square aperture. 
             Labels marked with an asterisk represent
             stellar CO absorption bandheads. The effect of dilution on these 
             bandheads is clearly seen.}
    \label{spectrumcircinus}
    \end{center}
\end{figure*} 

In Figure~\ref{spectrumcircinus} we present a spectrum of Circinus at
a spectral resolution of $R\sim4200$,
integrated  over a 0.8\arcsec square aperture centered on the nucleus.
The most prominent of the emission lines are the coronal lines 
[Si{\sc vi}] at 1.96 $\mathrm{\mu m}$ and the  
[Ca{\sc viii}] at 2.32 $\mathrm{\mu m}$. 
We also detect the [Al{\sc ix}] 2.04
$\mathrm{\mu m}$ line, first mentioned by Maiolino et al. (\cite{maio98}). 
In addition to these high excitation lines, the $\mathrm{H_2}$ 1-0\,S(1) 
2.12 $\mathrm{\mu m}$ and 
$\mathrm{Br\gamma}$ 2.17 $\mathrm{\mu m}$ emission lines are clearly
recognizable. 
The line fluxes, which are presented in Table~\ref{fluxes}, are
comparable to previous measurements  
(Maiolino et al. \cite{maio98}).

The $K$-band flux density, calculated from the continuum image shown in 
Figure ~\ref{linemaps} within an aperture of 0.8\arcsec\ is 
$3.1\times10^{-14}$ $\mathrm{W m^{-2}}$ $\mathrm{\mu m}^{-1}$
(approximately 50\,mJy). 
The stellar features, traced by the CO bandheads, are also distinguishable 
in the spectrum. However, 
not all of the stellar features are diagnostically useful.
For example, the $^{12}$CO\,(3-1) band is partially filled by the 
[Ca{\sc viii}] emission line, and the rest of the bandheads with wavelengths 
longer than 2.34 $\mathrm{\mu m}$ are affected by residual atmospheric
features or other emission lines. Therefore, only the $^{12}$CO\,(2-0) 
band was useful to extract information about the fraction of the 
nuclear flux that is indeed stellar.

Images of the 2.2-$\mathrm{\mu m}$ continuum and the Br$\gamma$,
$\mathrm{H_2}$ (1-0)\,S(1), [Ca\,{\sc viii}], and [Si\,{\sc vi}] 
line emission, as well as the $^{12}$CO\,(2-0) bandhead flux are presented 
in Figure~\ref{linemaps}. The location of the nucleus, as defined by
the centroid of  
the continuum emission, has been marked with an encircled cross in all maps.
This source is offset by $\sim0.15\arcsec$ to the south-east of the
peak of the Br$\gamma$ and H$_2$ 1-0\,S(1) line emission as well as of the
stellar light. 
Such an offset is consistent with 
the results of Prieto et al. (\cite{pri04}) who found that the center of the
$K$-band radiation is located at $\sim0.15\arcsec$ south-east 
of the brightest central emission seen in the $J$-band and HST F814W
images.
In contrast to these authors who argued that the $K$-band
source is 
fully obscured at shorter wavelengths and that the emission at these
shorter wavelengths is due to nuclear 
light scattered by the compact dusty structure surrounding the
Circinus nucleus, our results suggest a different
interpretation because the offset remains even in the $K$-band but only
for particular components of the line and continuum emission.
We argue that, at least for near infrared emission, the offset arises
because the star formation is centered $\sim$0.15\arcsec\ to the north
west of the non-stellar continuum peak.
The issue of extinction is addressed further in Section~\ref{density}.

\section{Nuclear dust emission} \label{dustemission}

\begin{figure*}
    \begin{center}
     \resizebox{0.9\textwidth}{!}{\includegraphics[clip, angle=-90]{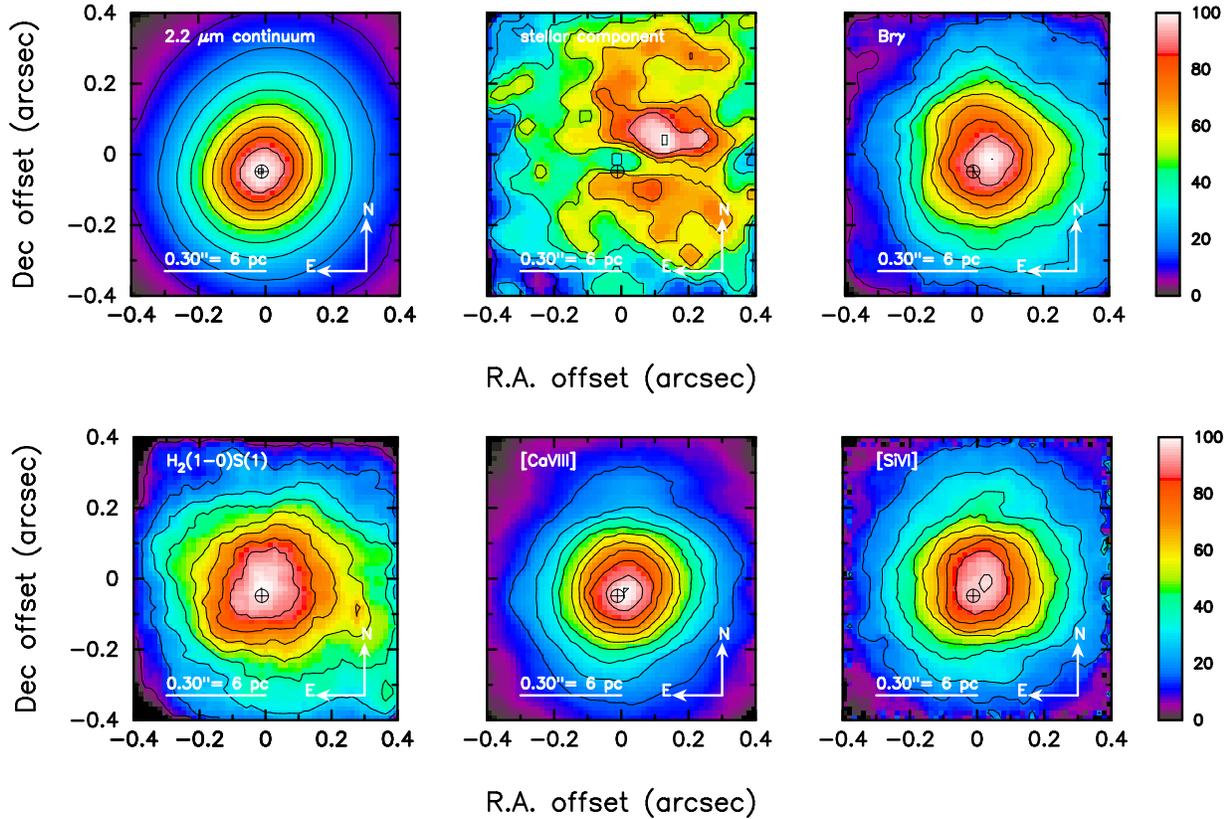}}    
    \caption{Intensity images extracted from the SINFONI data cube in the central arcsec 
             of Circinus. In each case, the colour scale extends from 0-100\% 
             of the peak flux, and contours are spaced equally between 20\% and 90\% 
             of the peak flux. An encircled cross indicates in each case the peak of 
             the continuum emission. The maps show, starting from the left corner,
             \textit{Top left}: 2.2$\mathrm{\mu m}$ continuum, \textit{Top center}: 
             stellar component 
             (flux contained in the stellar absorption bandhead $^{12}$CO(2-0)),
             \textit{Top right}: Br$\gamma$,  
             \textit{Bottom left}: H$_2$ 1-0\,S(1),
             \textit{Bottom center}: [Ca\,{\sc viii}], 
             and \textit{Bottom right}: [Si\,{\sc vi}].}
    \label{linemaps}
    \end{center}
\end{figure*} 

\begin{figure}
    \begin{center}
     \resizebox{\hsize}{!}{\includegraphics[clip]{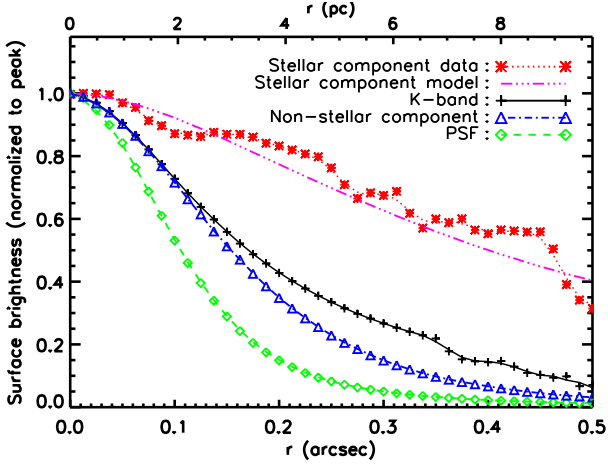}}   
    \caption{Radial profile of the nuclear $K$-band surface brightness
             obtained from  
             an integrated image of the SINFONI datacube 
             (crosses connected by a solid line).
             The SINFONI $K$-band stellar component (same as stellar continuum) 
             is represented by asterisks connected by a dotted line.
             The triple-dotted-dashed line represents a model fitted 
             to the stellar component data, 
             which is the exponential profile convolved with the PSF. 
             Open triangles connected by a dotted-dashed line indicate 
             the fraction of the $K$-band radiation which does not come from stars.
             The open diamonds connected by a short-dashed line indicate the PSF 
             profile.}
    \label{profiles1}
    \end{center}
\end{figure}

The observed equivalent width of the $^{12}$CO\,(2-0) bandhead was
used to obtain  
the fraction of the nuclear flux from the stars in this region, 
by assuming that the intrinsic (i.e. purely stellar) equivalent width in the
nuclear region must be consistent with values predicted by the models
created with the stellar population synthesis code STARS
(\cite{ste98,tho00}, Davies et al. \cite{dav03}).
We used the measured flux of the $^{12}$CO\,(2-0) band to estimate the
amount of non-stellar continuum using the equation: 
\[
%\begin{equation}
        1-D = \frac{EW_{\mathrm{obs}}}{EW_{\mathrm{intr}}} 
%\end{equation}
\]
where $D$ is the dilution factor, that is the ratio of stellar to
total continuum, 
and $EW_{\mathrm{obs}}$ and $EW_{\mathrm{intr}}$ are the observed and
intrinsic stellar equivalent widths respectively. 
The intrinsic equivalent width is 
nearly independent of star formation history for any ensemble of stars
as predicted by the STARS models, having an almost constant value of
$\sim12$\,\AA\ assuming solar metallicity (Davies et al. \cite{davies06}).
The observed stellar equivalent width in the central 0.8\arcsec\ was
found to be $\sim$1.8\,\AA, implying that 
$\sim$85\% of the total nuclear 
$K$-band luminosity in this region originates from the non-stellar
component, a fraction that increases on smaller scales. 
In Figure~\ref{profiles1}, the $K$-band continuum azimuthally averaged
radial surface brightness profile is plotted along with those of the
PSF and stellar and non-stellar components. 
The stellar component represents the radial profile of the 
stellar absorption bandhead $^{12}$CO(2-0) image in Fig.~\ref{linemaps}.
All profiles are normalized to their respective peak values;
and the stellar profile is centered 0.15\arcsec\ north west of the
non-stellar profile.
The predominance of the non-stellar component on these scales is
vividly demonstrated by the similarity of the non-stellar and total
$K$-band profiles, as well as by the red spectral slope in
Fig.~\ref{spectrumcircinus}.

Such nuclear $K$-band emission is interpreted as thermal radiation
from hot dust which is heated by the intense UV and X-ray radiation
emitted by the Seyfert nucleus.
It is often assumed that the dust is close to the sublimation limit
($\sim1600$\,K) since it is the
hottest emission which will have the greatest impact in the near
infrared.
However, the slope of the nuclear spectrum in the $K$-band indicates 
a temperature (unreddened) closer to 740\,K, 
similar to that found for NGC\,1068 by
Thatte et al. (\cite{thatte97}), but significantly more than the 300\,K
derived by Prieto et al. (\cite{pri04}) from modelling the broad band colours.
Since the spectral range of the $K$-band is smaller than 
the 1--10$\mu$m interval used by Prieto et al. (\cite{pri04}), 
the spectral slopes in these ranges are different,
but nevertheless compatible. 
The estimated temperature of $\sim$700\,K from the $K$-band continuum 
characterizes only the hottest dust, 
while 300\,K dust would dominate the mid-infrared SED.

As can be seen from Figure~\ref{profiles1}, the non-stellar $K$-band source 
is marginally resolved.
Indeed, a quadrature correction of its FWHM ($\sim 0.3 \arcsec$) with
that of the PSF 
yields an intrinsic size of $\sim 0.2 \arcsec$ (3.9 pc). 
The factor of 2 discrepancy with the size reported
by Prieto et al. (\cite{pri04}) arises primarily from the spatial
resolution estimated in the NACO data. Compared to their
0.16\arcsec\ our estimate is 0.14\arcsec, which would give a source
size of $\sim0.15\arcsec$.
This emphasises the difficulty of estimating the intrinsic size of
only marginally resolved sources.
Within the uncertainty of about 0.05\arcsec, the sizes are consistent 
with each other and with those predicted by the torus models 
(Pier \& Krolik \cite{pierandkrolik}, \cite{pie93}; 
Nenkova et al. \cite{nenkova})
in which the inner edge of the dust
torus lies $\sim$1 pc away from the AGN.

\section{Nuclear star formation} \label{starformation}

\begin{figure}
    \begin{center}
     \resizebox{\hsize}{!}{\includegraphics[clip]{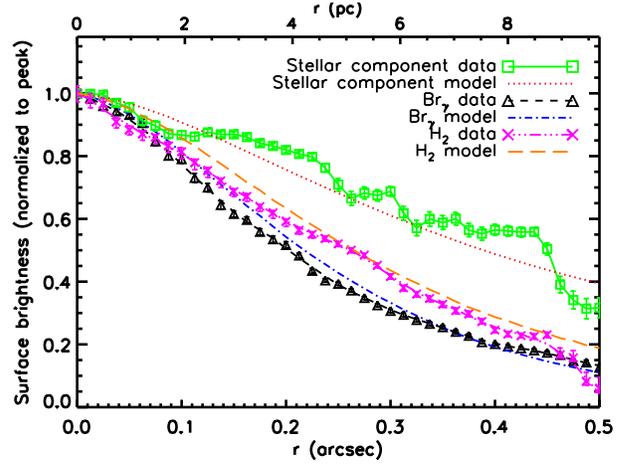}}   
    \caption{Radial profile of the surface brightness of the H$_2$ 1-0\,S(1) 
            (crosses connected by a triple-dotted-dashed line) emission line. 
            The long-dashed line represents a model fitted 
            to the linemap, which is the exponential profile convolved with the PSF.
            The narrow Br$\gamma$ radial profile is represented by open triangles 
            connected by a short-dashed line and its model by the dotted-dashed line.
            Open squares connected by a solid line indicate the nuclear stellar profile,
            whereas its model is represented by the dotted line.} 
    \label{profiles2}
    \end{center}
\end{figure}

Images of the Br$\gamma$ and H$_2$ 1-0\,S(1) line emission, which are
often associated with star formation, are presented in Figure~\ref{linemaps}. 
As the AGN is highly obscured (see Oliva et al. \cite{oliva98} 
and Matt et al. \cite{matt96}, \cite{matt99}), no broad line region
is visible. 
Several lines of evidence suggest that the Br$\gamma$ emission is
associated with star formation activity surrounding the Seyfert
nucleus rather than the narrow line region and ionisation cone. 
These are the similarity of the morphologies of the
Br$\gamma$, $^{12}$CO\,(2-0), and H$_2$ 1-0\,S(1) emission: 
the $^{12}$CO\,(2-0) is only slightly more extended than the two lines,
and all three are offset to the north west.
Additional support is provided by the consistency and uniformity of
the Br$\gamma$ and H$_2$1-0\,S(1) velocity fields and dispersion maps (see
Section~\ref{kinematics} and Figure~\ref{velmaps}).
In particular, the velocity field of the Br$\gamma$ has a gradient
matching that of the galaxy's major axis rather than showing signs of
outflow along the minor axis.
We note in fact that the H$_2$ 1-0\,S(1) appears to be slightly more extended
than the Br$\gamma$, particularly at weaker
levels westwards in the direction of the ionization cone (Maiolino et
al. \cite{maio98}); 
and it is only in this region that there are significant differences
between their velocity fields, with the 1-0\,S(1) showing signs of
outflow. 
Figure~\ref{profiles2} shows the azimuthally averaged surface
brightness radial profiles of the H$_2$ 1-0\,S(1) and narrow
Br$\gamma$ emission lines and the $^{12}$CO\,(2-0) stellar absorption
bandhead. 
%This Figure show that although the Br$\gamma$ is the less extended 
%of the three profiles, it is almost as extended as the H$_2$ 1-0\,S(1) 
%emission and much more extended to what it would be expected 
%if it was excited by the AGN ().   

\begin{figure*}
    \begin{center}
     \resizebox{0.9\textwidth}{!}{\includegraphics[clip, angle=-90]{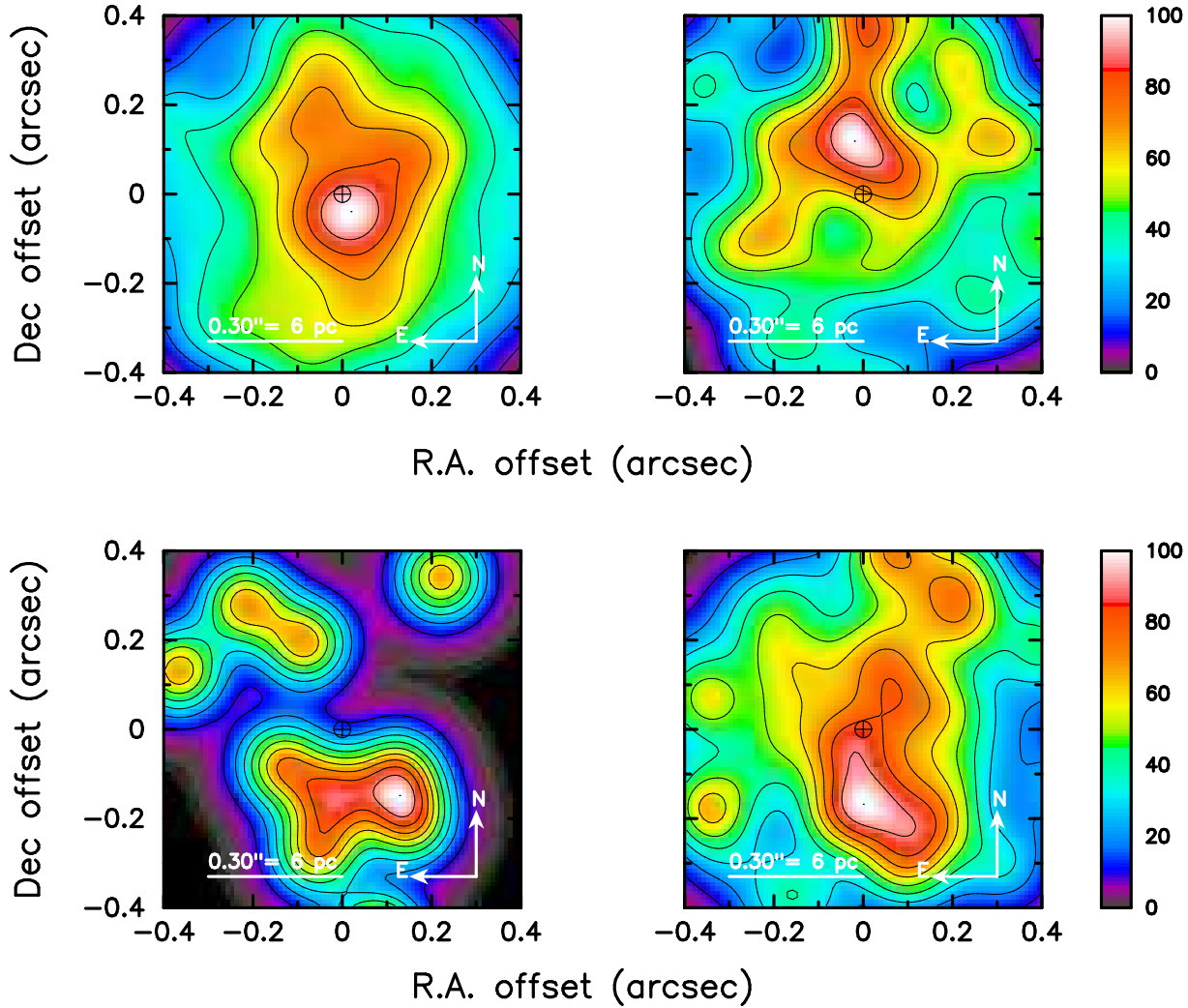}}    
    \caption{Intensity images of the star clusters simulations in the central arcsec 
             of Circinus. In each case, the colour scale extends from 0-100\% 
             of the peak flux, and contours are spaced equally between 10\% and 90\% 
             of the peak flux. An encircled cross indicates in each case the center of 
             the region. The maps show, starting from the left corner,
             \textit{Top left}: 1000 star clusters with constant mass 
	     $1\times10^3$ $M_{\sun}$, 
             \textit{Top right}: 100 star clusters with constant mass 
	     $1\times10^4$ $M_{\sun}$, 
             \textit{Bottom left}: 10 star clusters with constant mass 
	     $1\times10^5$ $M_{\sun}$, and 
	     \textit{Bottom right}: 300 star clusters with masses varying 
	     from $1\times10^3$ $M_{\sun}$ to $5\times10^4$ $M_{\sun}$. }
    \label{SC_simulations}
    \end{center}
\end{figure*} 

Both an exponential profile and $r^{1/4}$ de Vaucouleurs profile
provide good matches to each of these. 
We prefer to parameterise the profile with the former,
because of the clear evidence of a circumnuclear disk from
the existence of certain features (namely a bar, Maiolino et
al. \cite{maio00}; and a ring, Marconi et al. \cite{marconi94b}).
At these scales, less than 20 pc, the Br$\gamma$, and H$_2$ 1-0\,S(1)
data are characterised by a
disk-scale length $r_{\mathrm{d}}$ = $4^{+0.1}_{-0.2}$ pc.
Due to the symmetry in the morphology of the Br$\gamma$, and H$_2$ 1-0\,S(1) 
profiles, the profiles were not corrected for inclination. 
This would perhaps be a surprising result for an inclined disk, but indeed
the kinematics and simulations suggest that
the thickness of the disk plays an important part in the observed
morphology: a disk inclined at 65$^\circ$ with a thickness of only
4\,pc FWHM -- equivalent to an exponential scale height of 1.7\,pc --
is consistent with the symmetrical morphology.
We return to this point in Section~\ref{kinematics} where we discuss
the kinematics in more detail. 
%\textbf{Since the line emission is well matched by an exponential
%  profile, we have similarly fitted the stellar component to provide
%  a comparable characterization of what is a complex 
%  morphology.}

The complex morphology of the stellar component was analyzed in more detail 
by several simulations. Since the observed stellar luminosity is
expected to be produced by several clusters, simulations were carried out in order to 
study the way the clusters superpose to create the observed complex morphology. 
In the simulations each cluster was defined by two parameters: its position 
inside the field, and its mass. The positions were assigned randomly
following the radial distribution of the stellar component 
in Fig.~\ref{profiles2}, so that the
probability of finding a cluster in any position depends on an
exponential profile slightly more extended than the line emission.
For the mass distribution, systems of young clusters, including super star clusters 
which are preferentially found at the very heart of starbursts, 
appear to be well represented by a power law mass and luminosity function with 
$\alpha=-2$. 
The range of masses was from $1\times10^3$ to $3\times10^5$ $M_{\sun}$ 
which are typical values for star clusters (Meurer et al. \cite{meurer95}). 
Each cluster was characterized by a two dimensional Gaussian function with 
FWHM = 0.2$\arcsec$, our spatial resolution. 
The clusters were generated randomly over a circular field of 0.5$\arcsec$ radius. 
According to the mass distribution, a mass was assigned to each of them 
until the observed stellar mass in the region was reached, 
namely $1\times10^6$ $M_{\sun}$ (see below). 
Each cluster had the same mass-to-light ratio.
The summation of all clusters over the whole field yields the
overall morphology of the stellar component.
Fig.~\ref{SC_simulations} shows the results of the simulations. 
It can be seen from Fig.~\ref{SC_simulations} that when massive star clusters are 
present in the field, the clumpiness of the whole set increases. 
On the other hand, for a low mass distribution the field 
remains uniform and symmetric, and therefore the clumpiness decreases. 

The observed morphology in Fig.~\ref{SC_simulations} can be best reproduced by a
set of star clusters with a mass distribution ranging from 
$1\times10^3$ to $5\times10^4$ $M_{\sun}$, or alternatively a constant
mass with value of $1\times10^4$ $M_{\sun}$, although other scenarios
cannot be excluded .
Our simulations indicate that the observed complexity and asymmetry of the
morphology of the stellar component could easily arise from the
superposition of a number of star clusters with typical mass $10^4$
$M_{\sun}$.

\begin{figure*}
    \begin{center}
     \resizebox{0.4\textwidth}{!}{\includegraphics[clip]{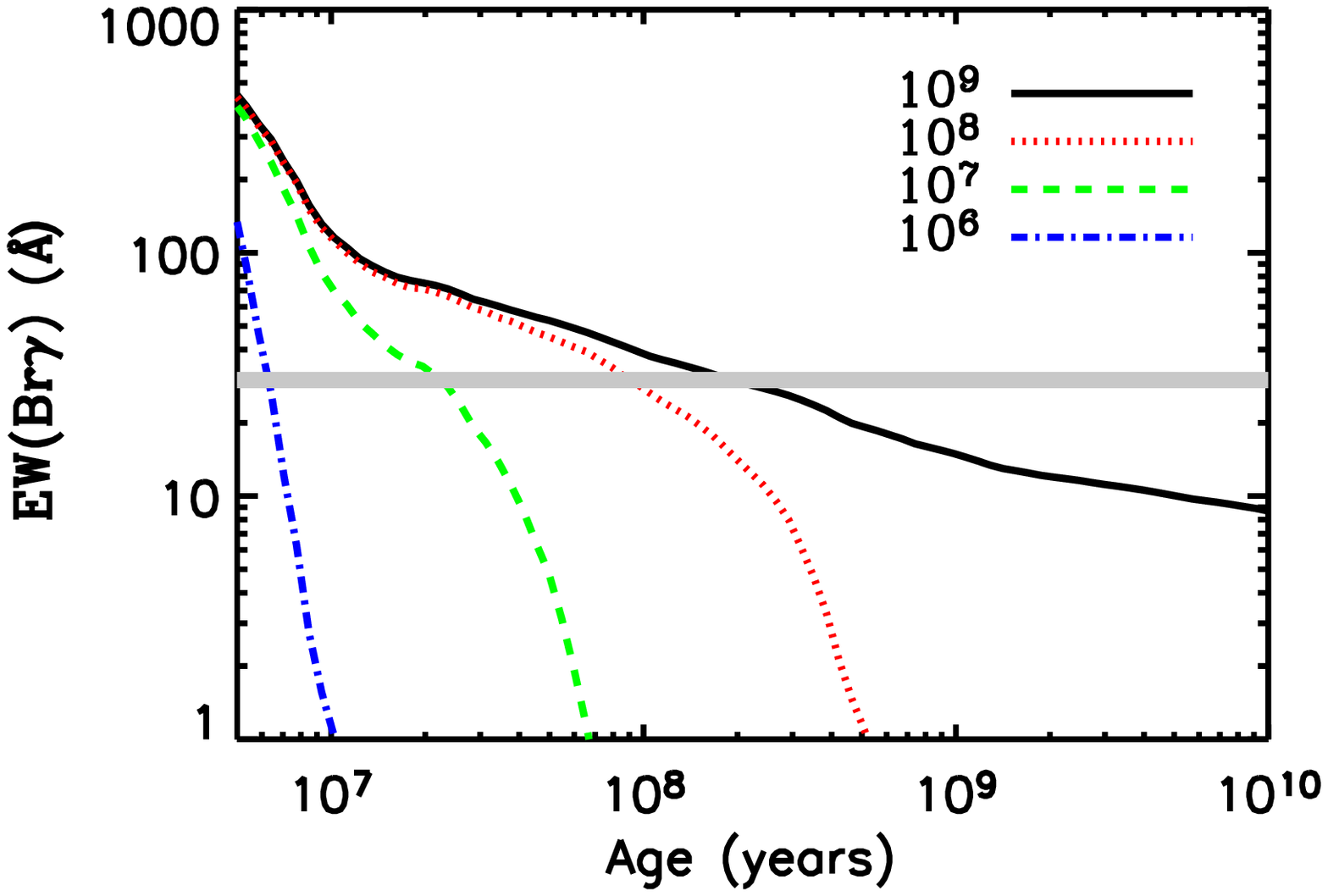}}   
     \hspace{0.5cm}
     \resizebox{0.4\textwidth}{!}{\includegraphics[clip]{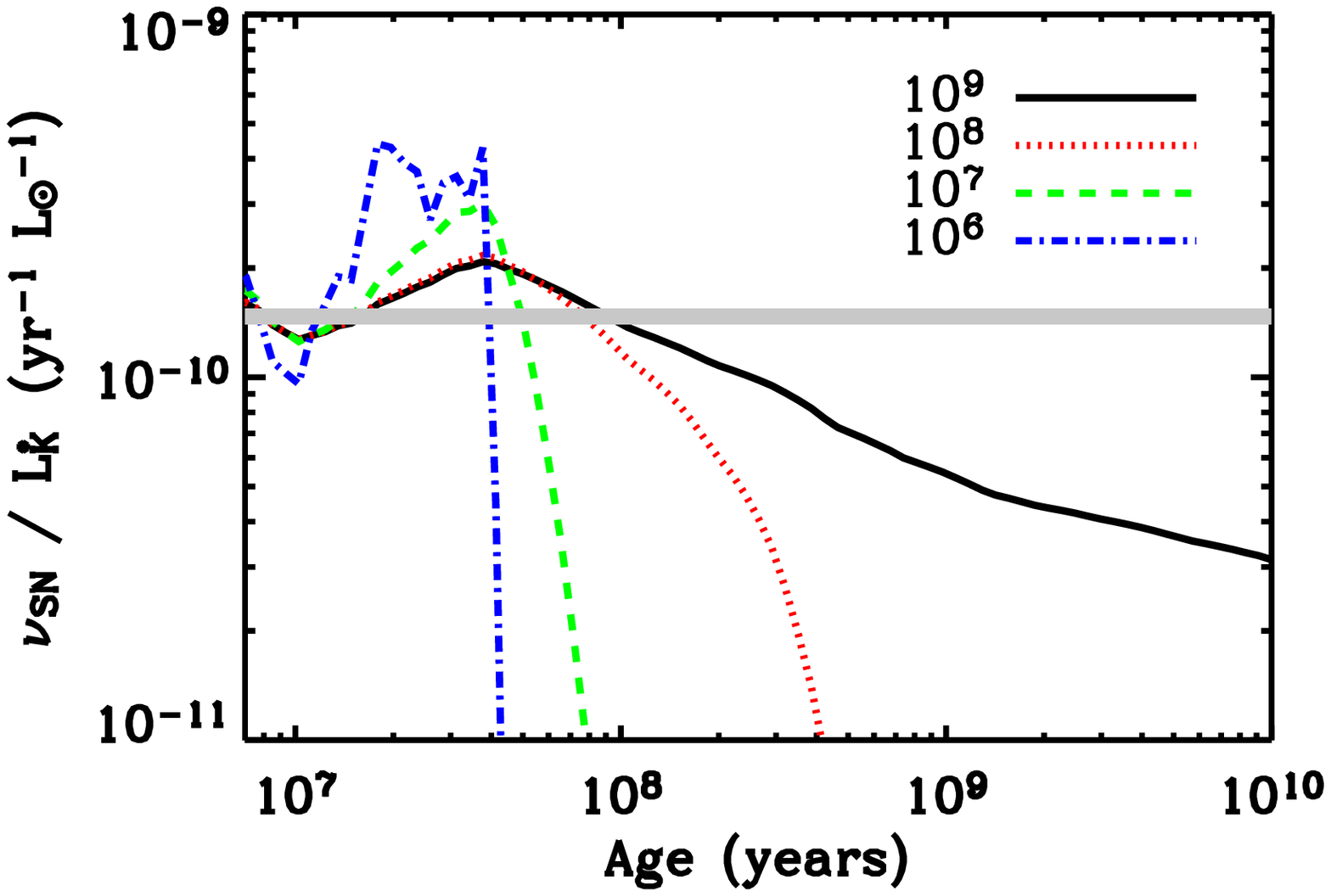}}
     \vspace{0.5cm}
     \resizebox{0.4\textwidth}{!}{\includegraphics[clip]{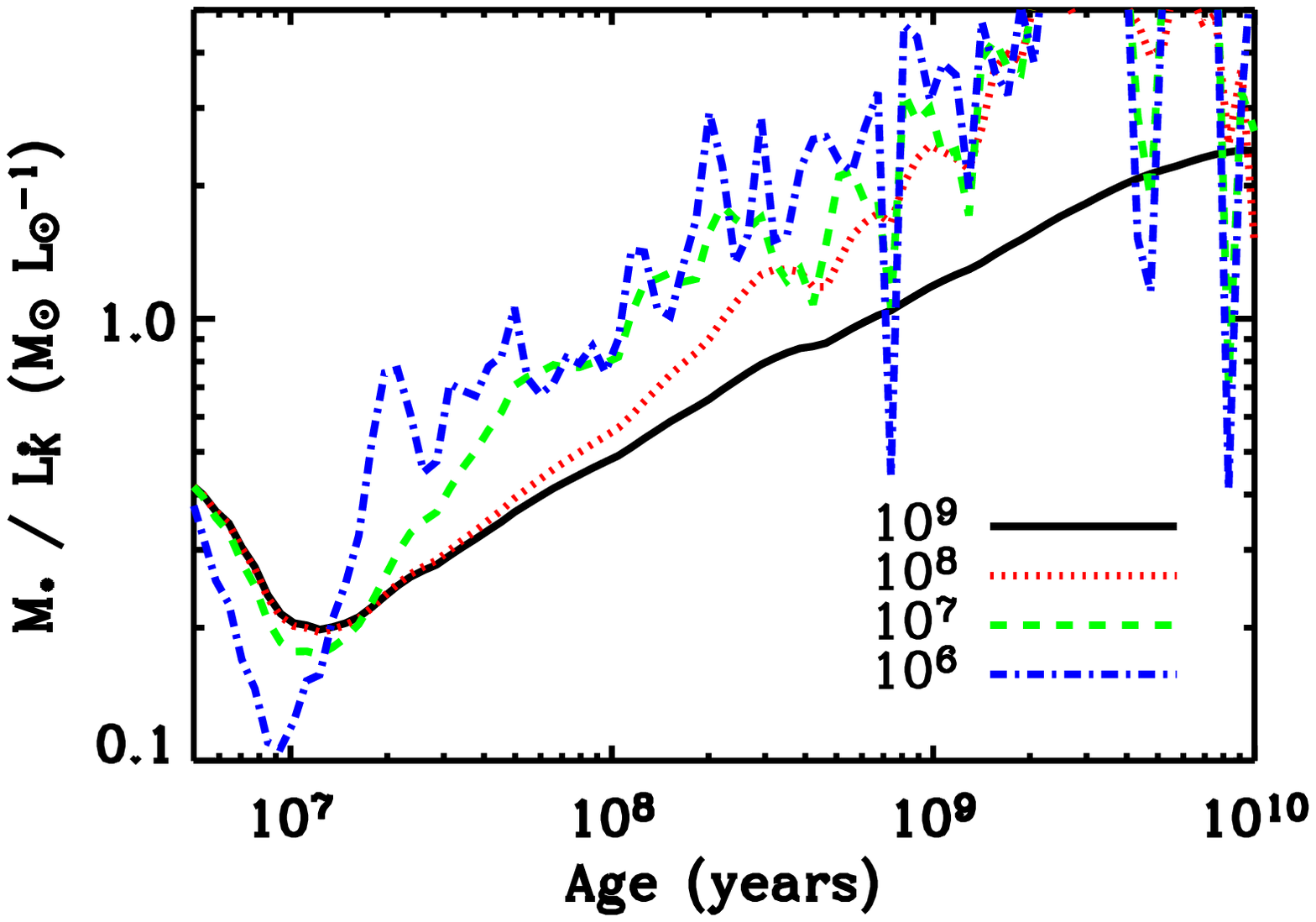}}
     \hspace{0.5cm}  
     \resizebox{0.4\textwidth}{!}{\includegraphics[clip]{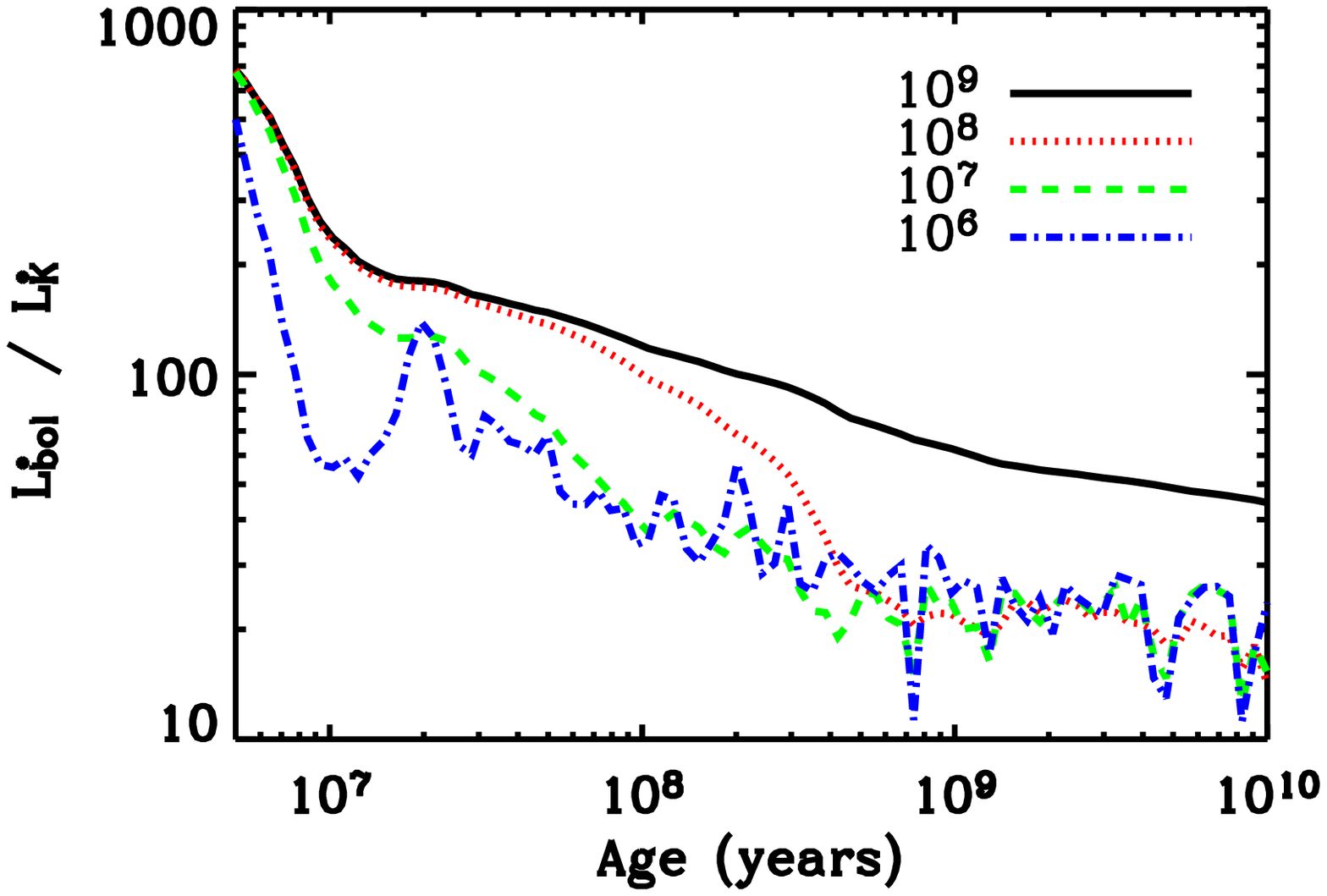}}   

    \caption{Evolution of $EW(\mathrm{Br\gamma})$ (\textit{Top left}), 
             $\nu_{\mathrm{SN}}$/$L_{\mathrm{K}}^*$ (\textit{Top right}), 
             $M_*$/$L_{\mathrm{K}}^*$ (\textit{Bottom left}) and 
             $L_{\mathrm{bol}}^*$/$L_{\mathrm{K}}^*$ (\textit{Bottom right}) with age for 
             a stellar cluster with four star formation histories. 
             The gray bands indicate the range of the observed ratios. 
             An asterisk indicates that the parameter refers
             to its stellar component.}
    \label{starburstmodels}
    \end{center}
\end{figure*}

We determine the age of the star formation by means of two diagnostics: 
the intrinsic equivalent width of the Br$\gamma$ line and the
supernova rate to stellar 
$K$-band luminosity ratio ($\nu_{\mathrm{SN}}$/$L_{\mathrm{K}}^*$) in
the nuclear region. 
These provide independent information on the star formation history, 
the former representing the number of hot young blue supergiant stars
and the latter the number of supernova remnants, both normalised to
the number of later type red giant and supergiant stars.
By comparing the two diagnostics to models, 
the age of the starburst can be satisfactorily constrained. 
The $EW(\mathrm{Br\gamma})$ was estimated from the observed Br$\gamma$
flux and the stellar continuum (i.e. corrected for dilution by
subtracting the continuum contribution from hot dust), resulting in
$EW(\mathrm{Br\gamma})$ = 30 \AA\ in a $0.8\arcsec$ aperture. 
The supernova rate $\nu_{\mathrm{SN}}$ was obtained using the
empirical relation presented in  
Condon (\cite{condon92}) with a spectral index $\alpha = 0.1$ 
(as measured between 5 and 8.64 GHz), and a 
flux density at 8.64 GHz of $S_{10}$ = 12 mJy, resulting in a value of 
$\nu_{\mathrm{SN}}$ = 2.21$\times 10^{-4}$ yr$^{-1}$.
The radio flux density was calculated by means of the 1\arcsec\
resolution radio image from Davies et al. (\cite{davies98}). 
The flux density at 8.64 GHz was summed over an aperture of
$0.8\arcsec$, to be consistent with our fluxes, giving 14 mJy. 
Then the thermal contribution  
was calculated as given in Condon (\cite{condon92}) 
for a standard electron temperature of 1$\times 10^4$ K. 
The resulting 2 mJy was subtracted from the total flux density at 8.64 GHz.
Dividing $\nu_{\mathrm{SN}}$ found in this way by the stellar 
$K$-band luminosity $L_{\mathrm{K}}^*$ of $1.5\times 10^6$ $L_{\sun}$ 
found after multiplying 
the observed $K$-band luminosity $L_{\mathrm{K}}$ by the dilution factor $D$,
yielded a ratio of 1.5$\times 10^{-10}$ $L_{\sun}^{-1}$ yr$^{-1}$.

We have used the evolutionary synthesis code STARS
(\cite{ste98,tho00}, Davies et al. \cite{dav03}) to determine the star formation history
assuming an exponentially decaying star formation rate of the 
form $exp(-t/t_{scl})$, 
where $t_{scl}$ is the burst decay time scale and $t$ is the age of
the star cluster.
We make the standard assumptions of solar metallicity, and a Salpeter IMF 
of slope  $\alpha = -2.35$ in the range 1--120 $M_{\sun}$.
An important advantage of this model is that it allows us to study not only
instantaneous and continuous scenarios, but also finite star formation
timescales.

%First, we compared the calculated intrinsic Br$\gamma$ equivalent
%width with that from STARS. 
%The latter is calculated from the 
%ratio between the Lyman continuum luminosity of the star cluster, arising
%from H {\sc ii} regions around hot young stars, 
%and the $K$-band continuum luminosity, which is mostly late-type giant
%and supergiant stars.
%We used the definition
%\[
%\begin{equation}
%        L_{\mathrm{K}}^*(L_{\sun}) = 
%              1.16\times10^4\times D(\mathrm{Mpc})^2\times S_{\mathrm{K}}^*(\mathrm{mJy})  
%\end{equation}
%\]
%
%given by Genzel et al. (\cite{genzel95}) with a $K$-band bandwidth of
%0.6\,$\mu$m to obtain the $K$-band flux density. 
%Using the model $L_{\mathrm{Lyc}}$ luminosity, 
%the Br$\gamma$ flux can be estimated via
%\[
%\begin{equation}
%        L_{\mathrm{Lyc}}(L_{\sun}) = 
%              5.37\times10^{19}\times F_{\mathrm{Br\gamma}}
%              (\mathrm{ergs^{-1} cm^{-2}})
%              \times D(\mathrm{Mpc})^2  
%\]
%\end{equation}
%
%Using these equations, one finds that model Br$\gamma$ equivalent width
%is simply given by the ratio
%\[
%\begin{equation}
%EW(\mathrm{Br\gamma}) = 
%         3.4\ \frac{L_{\mathrm{Lyc}}(L_{\sun})}{L_{\mathrm{K}}^*(L_{\sun})}
%\]
%\end{equation}
%

We obtain different ages for the starburst depending on the timescale of 
the star formation.
We consider 4 timescales:
continuous star formation with $t_{scl} = 10^{9}$\,yr, 
an instantaneous burst with $t_{scl} = 10^{6}$\,yr, and two
intermediate scenarios with $t_{scl} = 10^{7}$ and $10^8$\,yr.
First, we compared the calculated intrinsic Br$\gamma$ equivalent
width with that from STARS.
Figure~\ref{starburstmodels} shows the evolution of $EW(\mathrm{Br\gamma})$,
$\nu_{\mathrm{SN}}$/$L_{\mathrm{K}}^*$, $M_*$/$L_{\mathrm{K}}^*$ and 
$L_{\mathrm{bol}}^*$/$L_{\mathrm{K}}^*$ 
as functions of cluster age for each model. 
It can be seen from this figure that $EW(\mathrm{Br\gamma})$ alone
cannot constrain the age since all scenarios are plausible.
%Therefore, another constraint was needed. 

%A first approach was made by estimating the mass of the stars 
%within the 0.8\arcsec\ aperture for each scenario and comparing it
%with the dynamical mass (see Section~\ref{kinematics}), which 
%establishes an upper limit for 
%the total mass contained in the region.
%A model can be ruled out if it gives a mass bigger than the dynamical mass. 
%However, we obtain for each scenario a mass of the stars rather
%lower than the dynamical mass, so that the age of the starburst can
%not be constrained in this way.
%On the other hand, by 
Using the ratio
$\nu_{\mathrm{SN}}$/$L_{\mathrm{K}}^*$ together with $EW(\mathrm{Br\gamma})$, 
we note that within the central $R<8$ pc, the constant and 
the instantaneous star formation models do not fit completely both of 
the observational constraints ($EW(\mathrm{Br\gamma})$ = $30$\,\AA\ and 
$\nu_{\mathrm{SN}}$/$L_{\mathrm{K}}^*$ = $1.47\times10^{-10}$
$L_{\sun}^{-1}$ yr$^{-1}$). 

The star formation timescales $t_{scl} = 10^7$ and $t_{scl} = 10^8$\,yr 
do fit these constrains from the intersections points at $2\times 10^7$ 
and $8\times 10^7$\,yr respectively.
While the data are consistent with these two scenarios, the shorter
timescale would require a degree of fine-tuning, 
particularly since the supernovae responsible for the radio conntinuum are only 
just beginning to appear and $EW(\mathrm{Br\gamma})$ is decaying relatively quickly. 
Thus there is only a small range of possible ages for this timescale.
At an age of 80\,Myr, the constraints give rise to a wider range of
allowed ages, suggesting this is the more likely scenario.
Crucially, in either case the data indicate the presence of a young stellar
population within a few parsecs of the active nucleus which is now
less than half as intense as when it began.

For the preferred star formation scenario given above we find a current 
mass to $K$-band luminosity ratio of $\sim$0.6\,$M_\odot/L_\odot$,
yielding a stellar mass of $M_*$ $\approx$ $1\times 10^6$ $M_{\sun}$. 
Another important parameter to analyze 
is the bolometric luminosity attributable to the nuclear star formation. 
Figure~\ref{starburstmodels} shows the evolution of the bolometric
luminosity to $K$-band luminosity ratio for an ensemble of stars. 
From this Figure we obtain $L_{\mathrm{bol}}^*$/$L_{\mathrm{K}}^*$ $\approx$ 150. 
By multiplying this ratio by the observed stellar $K$-band luminosity of
$L_{\mathrm{K}}^* = 1.5\times 10^6$ $L_{\sun}$, 
we find that the star formation in this region has 
$L_{\mathrm{bol}}^*$ = $2.3\times 10^8$ $L_{\sun}$ and accounts for $1.4\%$ 
of the bolometric luminosity of the entire galaxy 
($L_{\mathrm{bol}}$ = $1.7\times 10^{10}$ $L_{\sun}$, Maiolino et al. \cite{maio98}).  

Let us now assume a screen extinction of $A_{\mathrm{V}}$ = 9 mag, as
suggested by Maiolino et al. (\cite{maio98}) for the nuclear region. 
Since $A_{\mathrm{K}}$ $\approx$ $A_{\mathrm{V}} / 10$ 
(Howarth \cite{howarth83}),
$L_{\mathrm{K}}^*$ is $\sim$2 times that observed. 
This consideration has no effect 
on the $EW(\mathrm{Br\gamma})$, but it reduces the 
$\nu_{\mathrm{SN}}$/$L_{\mathrm{K}}^*$ by a factor of two. 
This new constraint does not change the prefered star formation scenario,  
still suggesting that the star formation is very young. 
The bolometric luminosity attributable to the nuclear star formation
in this case would be $L_{\mathrm{bol}}^*$ = $4.5\times 10^8$\,$L_{\sun}$.

A perhaps more physical model than screen extinction is mixed
extinction, where the dust and gas are uniformly mixed with the stars.
In either case, the near infrared color excess is defined as 
%\[
%E_{H-K} = (H-K) \ - \ (H_0-K_0)
%\]
%where $H$ and $K$ are the observed magnitudes, and $H_0$ and $K_0$ are
%the intrinsic magnitudes. This can be written as
\[
%\begin{equation}
        E_{H-K} = -2.5\log\bigg[\frac{I_H/I_{0H}}{I_K/I_{0K}}\bigg] 
%\end{equation}
\]
where, $I_H$ and $I_K$ are the observed intensities in the
two bands, and $I_{0H}$ and $I_{0K}$ are the intrinsic intensities.
One can then use either $I_\lambda/I_{0\lambda} = e^{-\tau_\lambda}$
for the screen model or 
$I_\lambda/I_{0\lambda} = (1-e^{-\tau_\lambda})/\tau_\lambda$ for
the mixed model to derive the optical depth.
With the usual wavelength scaling that 
$\tau_{\rm V} = 5.5\tau_{\rm H} = 9.7\tau_{\rm K}$
one can then estimate the optical depth in the visual.
The near infrared excess $E_{H-K} = 0.45$--0.60 reported by Maiolino
et al. (\cite{maio98}) then implies an extinction, after accounting
for the galactic foreground, 
%screen of $A_{\rm V}=1.5$ which reduces
%the intrinsic excess to $E_{H-K} = 0.37$--0.52,
of $\tau_{\rm V} = 4.4$--6.2 for the screen model, and 
$\tau_{\rm V} = 12$--25 for the mixed model, similar to the estimate
of Prieto et al. (\cite{pri04}).
We note also that for the mixed model, the colour excess will
saturate because if $\tau_\lambda$ is greater than a few, 
$I_\lambda/I_{0\lambda} \sim 1/\tau_\lambda$ so that the excess becomes 
$E_{H-K} \sim -2.5\log{\tau_K/\tau_H}$.
The limiting value of $E_{H-K} = 0.6$ is consistent with the upper end of the
measured range, suggesting that greater extinctions cannot be ruled out.
We can, nevertheless, impose a very strong limit on the maximum
possible extinction:
the intrinsic luminosity from this nuclear star formation cannot exceed the
galaxy's bolometric luminosity 
($L_\mathrm{bol}^{*,nuc} < L_{\mathrm{bol}}$).
The lower limit, that 
$L_\mathrm{bol}^{*,nuc} > L_\mathrm{bol}/75$,
comes from the starburst model under the assumption of no
extinction. 
The actual value of $L_\mathrm{bol}^{*,nuc}$ is governed by the scaling of
the starburst model.
And this scaling depends directly on the ratio
between the intrinsic and observed fluxes in the waveband where it is
determined.
Since the ratio between the intrinsic $K$-band and bolometric
luminosities 
of the nuclear stellar component is fixed by the starburst model, 
it follows that the intrinsic $K$-band stellar luminosity must be less
than 75 times the observed $K$-band stellar luminosity.
Assuming a mixed model, one has 
$I_K/I_{0K} \sim 1/\tau_{\rm K}$, implying that
 $\tau_{\rm K} \sim 75$, equivalent to $\tau_{\rm V} \sim 750$. 
A more meaningful, although less robust, limit is found if one
excludes the circumnuclear star formation which Maiolino et
al. (\cite{maio98}) estimated to account for $\sim10^{10}$\,L$_\odot$.
In this case $L_{\mathrm{bol}}$/$L_{\mathrm{bol}}^{*,nuc} < 30$ and
the maximum optical depth would be of order 
$\tau_{\rm V} \sim 300$.
Although these limits are large, they are important and 
further discussion on the extinction will be presented in
Section~\ref{density}.

\section{Gas kinematics and mass in the nuclear region} \label{kinematics}

\begin{figure*}
    \begin{center}
     \resizebox{\hsize}{!}{\includegraphics[clip, angle=-90]{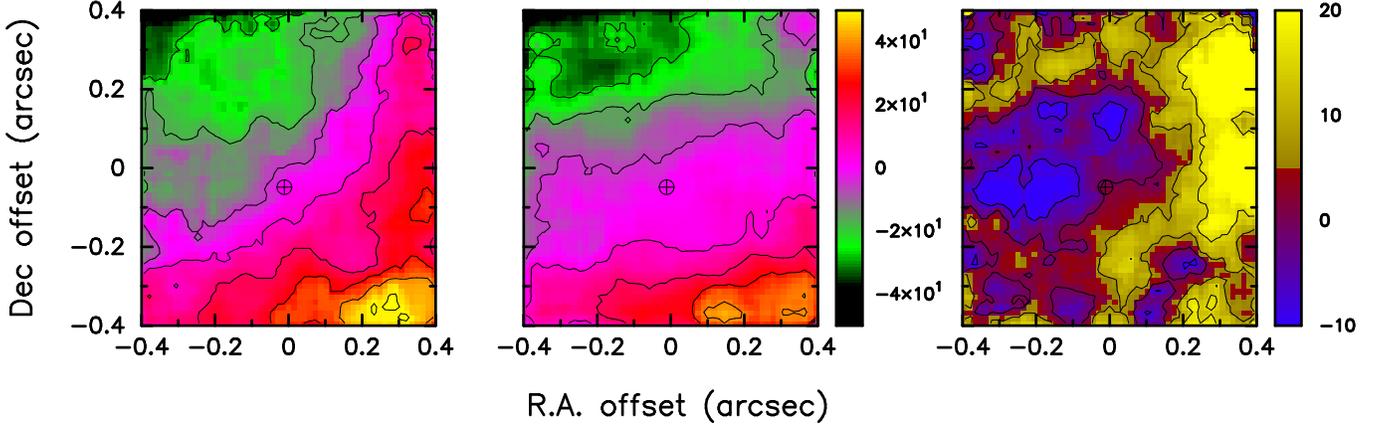}}   
    \caption{Velocity maps extracted from the SINFONI data cube in the central 
             $0.8\arcsec\times0.8\arcsec$ of Circinus. 
             The maps show, from left to right: 
             H$_2$ 1-0\,S(1), Br$\gamma$ and difference in the velocity fields
             of the two lines
             For the case of H$_2$ 1-0\,S(1) and Br$\gamma$ velocity maps, 
             the colour scale extends from [-50, 50] km s$^{-1}$, 
             and contours are spaced equally every 10 km s$^{-1}$.  
             For the difference in the velocity fields, the colour 
             scale extends from [-10, 20] km s$^{-1}$, 
             and contours are spaced equally every 5 km s$^{-1}$
             The difference in the velocity maps ranges between
             [-10,10] km s$^{-1}$, except for the western edge of the map, 
             where a difference of $\sim 25$ km s$^{-1}$ is observed.
             An encircled cross indicates in each case the peak of 
             the continuum emission.}
    \label{velmaps}
    \end{center}
\end{figure*}

In Figure~\ref{velmaps} the projected velocity maps of the H$_2$ and
the Br$\gamma$ lines are presented, as well as the difference 
in the velocity fields of the two lines. 
The maps for both emission lines are broadly consistent, 
exhibiting a gradual increase in velocity from north to south along 
a position angle (P.A.) of $\sim18\degr$ for the Br$\gamma$ and $\sim25\degr$ 
(excluding the righthand side, see below) for the 1-0\,S(1). 
Given the typical uncertainty of 5--10$\degr$ in each of these, they are consistent.
%Due to uncertainities in the data this 
%value can differ by $-7\degr$.
They are the same as the galaxy's major axis (Freeman et al. \cite{freeman}), 
a strong indication of pure galaxy rotation, and suggesting that there are
no warps in the galaxy down to scales at least as small as a few
parsecs. At smaller scales (from 0.1 to $\sim$0.4 pc), the maser 
emission traces a warped, edge-on accretion disk, which changes the
P.A. from $\sim$29$\degr$ at 0.1\,pc to $\sim$56$\degr$ at 0.4 pc
(Greenhill et al. \cite{greenhill03}).
Our results suggest that at larger scales ($\sim$1 pc), possibly at the 
starting point of the torus, the galaxy's position angle has reverted
back to $\sim$25$\degr$.
  
The fact that the P.A. for the H$_2$ 1-0\,S(1) velocity field appears
to be greater is
due primarily to the change in velocity gradient in a 0.3\arcsec\
strip along the western edge of the map;
excluding this region,
the difference between the maps is
statistically insignificant (see Figure~\ref{velmaps} right image). 
The western region coincides with the extension of emission in the
H$_2$ 1-0\,S(1) flux map and seems likely to be associated with a molecular
outflow in (or at the edge of) the ionisation cone, perhaps
radiatively driven by the AGN.  
It is important to notice that the observed velocity field of the Br$\gamma$ line 
is clearly different to the one expected 
if it was associated with the NLR of the AGN, which would exhibit an increase 
in velocity in the western direction along the minor axis of the galaxy, 
as an outflow in the ionisation cone.

By correcting the velocity maps for inclination ($i$ = 65$\degr$), 
we find a rotation velocity of 75\,km\,s$^{-1}$ at 8 pc from the nucleus. 
The velocity dispersion of the stars in the nuclear region is almost constant 
with a value of $\sigma_{\mathrm{obs}}$ $\approx$ 80\,km\,s$^{-1}$. 
By correcting the measured dispersion for the 
instrumental resolution corresponding to 
$\sigma_{\mathrm{inst}} \approx 32$ km $s^{-1}$,  
we obtain an intrinsic dispersion of $\sigma_{\mathrm{z}}$ $\approx$ 70\,km\,s$^{-1}$, 
yielding a velocity to dispersion ratio of
$V_{\mathrm{rot}}/\sigma_{\mathrm{z}}$ $\approx$ 1.1.
This borderline ratio suggests that while there is significant
rotation, random motions are also important -- and on these scales
perhaps dominant.
Indeed, the kinetic energy in random motions is of order $3\sigma^2$
while that in the ordered rotation is $V_{\mathrm{rot}}^2$.
This would imply that the distribution of gas is rather thick -- consistent
with our suggestion in Section~\ref{starformation} that the symmetry
of the observed isophotes is due to the physical height of the nuclear
region.

Further evidence that this is the case comes from the scale height
derived if one assumes that the gas and stars are self gravitating,
although this should only really be applied to a thin disk.
In this case
\[
%\begin{equation}
        \sigma_{\mathrm{z}}^2 = 2\pi \mathrm{G} \Sigma z_0 
%\end{equation}
\]
where $\sigma_{\mathrm{z}}$ is the velocity dispersion of the stars, 
$\Sigma$ is the surface density of the disk, and $z_0$ is 
a characteristic measure of the scale-height of the disk perpendicular to
the plane. In the nuclear region at $R<0.4\arcsec$ (8 pc), we estimate 
the mass of the gas and the stars residing in the disk to be $1.7\times 10^7$
$M_{\sun}$ (see also Section~\ref{density}), 
yielding a mean surface density of 
$\Sigma = 8.5\times 10^4$ $M_{\sun}$ pc$^{-2}$. 
By considering an intrinsic $\sigma_{\mathrm{z}}$ = 70 km/s, 
we obtain a height of $z_0 = 2$\,pc, consistent with our earlier
estimate based on the isophotal symmetry.

We suggest that this thickness does not necessarily imply that there
is a distinct spheroidal structure in the central few parsecs; but
that we are only looking at the very inner region of a larger disk
which extends out perhaps as far as circumnuclear ring.
By measuring the rotation velocity within a truncated field of view --
i.e. before $V_{\mathrm{rot}}$ reaches its asymptotic value -- 
one would underestimate $V_{\mathrm{rot}}/\sigma$.
It is also possible that either dynamics associated with the black
hole, or heating from the AGN or local intense star formation, have
caused the inner region of the disk to thicken.

For systems where the velocity dispersion is significant,
  estimating the mass from the rotational velocity alone leads to an
  underestimate. Accounting for random motions is not trivial, but can
  be approximated by 
$M_{\mathrm{dyn}}$ = $(V_{\mathrm{rot}}^2+3\sigma^2)r/G$. 
Hence we obtain a dynamical mass for the nucleus within
$\pm0.4\arcsec$ of $M_{\mathrm{dyn}}$ = $3.5\times 10^7$ $M_{\sun}$.

\section{Dimensioning the clumpy star forming torus in Circinus}
\label{density}

The spatial scales on which we have traced the gas and stars are
similar to those associated with standard torus models
(e.g. Pier \& Krolik \cite{pierandkrolik}, \cite{pie93}; 
Nenkova et al. \cite{nenkova}, Schartmann et al. \cite{sch05}, Fritz
et al. \cite{fri06}).
Indeed, the latter two authors have modelled Circinus specifically and
find size scale for the torus of 30\,pc and 12\,pc respectively.
However, to understand better the relation between the gas and the
stars, we need, in 
addition to the stellar mass found in Section~\ref{starformation}, to
estimate the gas mass.
Fortunately, as we discuss below, there is a way to do this from our data.

\begin{table*} 
        \begin{center}
        \begin{tabular}{l c c c c c c}
        \hline 
        \hline \noalign{\smallskip}
        Galaxy & log$\frac{L_{\mathrm{(1-0)S(1)}}}{L_{\sun}}^a$ & 
        log$\frac{M_{\mathrm{gas}}}{M_{\sun}}$ &
        log$\frac{L_{\mathrm{IR}}}{L_{\sun}}$ & 
        log$\frac{L_{\mathrm{(1-0)S(1)}}}{L_{\mathrm{IR}}}$ &
        log$\frac{L_{\mathrm{(1-0)S(1)}}}{M_{\mathrm{gas}}}$ & References$^b$ \\ 
        (1) & (2) & (3) & (4) & (5) & (6) & (7)\\
        \hline \noalign{\smallskip} 
        IR 01364-1042 & 6.7  & 10.2  & 11.8  & -5.0  & -3.4       & 1\\
        IR 05189-2524 & 7.0  & 10.4  & 12.1  & -5.1  & -3.4	  & 1         \\
        IR 09111-1007 & 6.9  & 10.4  & 11.9  & -5.0  & -3.4 	  & 1 \\ 
        IR 14378-3651 & 6.8  & 10.2  & 12.1  & -5.3  & -3.3	  & 1 \\
        IR 17208-0014 & 7.3  & 10.6  & 12.3  & -5.1  & -3.4	  & 1 \\ 
        NGC1614       & 6.2  &  9.7  & 11.6  & -5.3  & -3.5	  & 2 \\
        NGC2623       & 6.5  & 10.0  & 11.5  & -5.0  & -3.5	  & 2 \\
        IR 10173-0828 & 5.8  &  9.9  & 11.8  & -5.9  & -4.1	  & 2 \\
        NGC6090       & 6.7  & 10.5  & 11.5  & -5.0  & -4.0	  & 2 \\
        NGC6240       & 8.0  & 10.6  & 11.8  & -3.9  & -2.6	  & 2 \\
        NGC7469       & 6.4  & 10.2  & 11.6  & -5.2  & -3.8	  & 2 \\
        Mrk231        & 7.2  & 10.3  & 12.5  & -5.3  & -3.1	  & 2 \\
        Mrk273        & 7.3  & 10.6  & 12.1  & -4.8  & -3.3	  & 2 \\
        Arp220        & 7.0  & 10.5  & 12.2  & -5.2  & -3.5	  & 2 \\
        NGC695        & 6.4  & 10.6  & 11.7  & -5.2  & -4.2	  & 3 \\
        NGC1068       & 6.0  & 10.0  & 11.5  & -5.5  & -4.0	  & 3 \\
        NGC5135       & 6.2  & 10.2  & 11.1  & -4.9  & -4.0	  & 3 \\

        \hline
        \hline
        \end{tabular}
        \end{center}

$^a$ Reference for $L_{\mathrm{(1-0)S(1)}}$ data: 
  Goldader et al. (\cite{gold97}), 
  except for NGC\,6090 Sugai et al. (\cite{sug00}) \\
$^b$ References for $M_{\mathrm{gas}}$ data: 
     (1) Mirabel et al. (\cite{mirabel90}); 
     (2) Bryant \& Scoville (\cite{bryant99});
     (3) Gao \& Solomon (\cite{gao04}) \\

        \caption{Properties of the sample of luminous and ultraluminous 
                 infrared galaxies used to estimate 
                  $L_{\mathrm{(1-0)S(1)}}$/$M_{\mathrm{gas}}$}
\label{galaxies}
\end{table*}
\begin{figure*}
    \begin{center}
     \resizebox{0.4\textwidth}{!}{\includegraphics[clip]{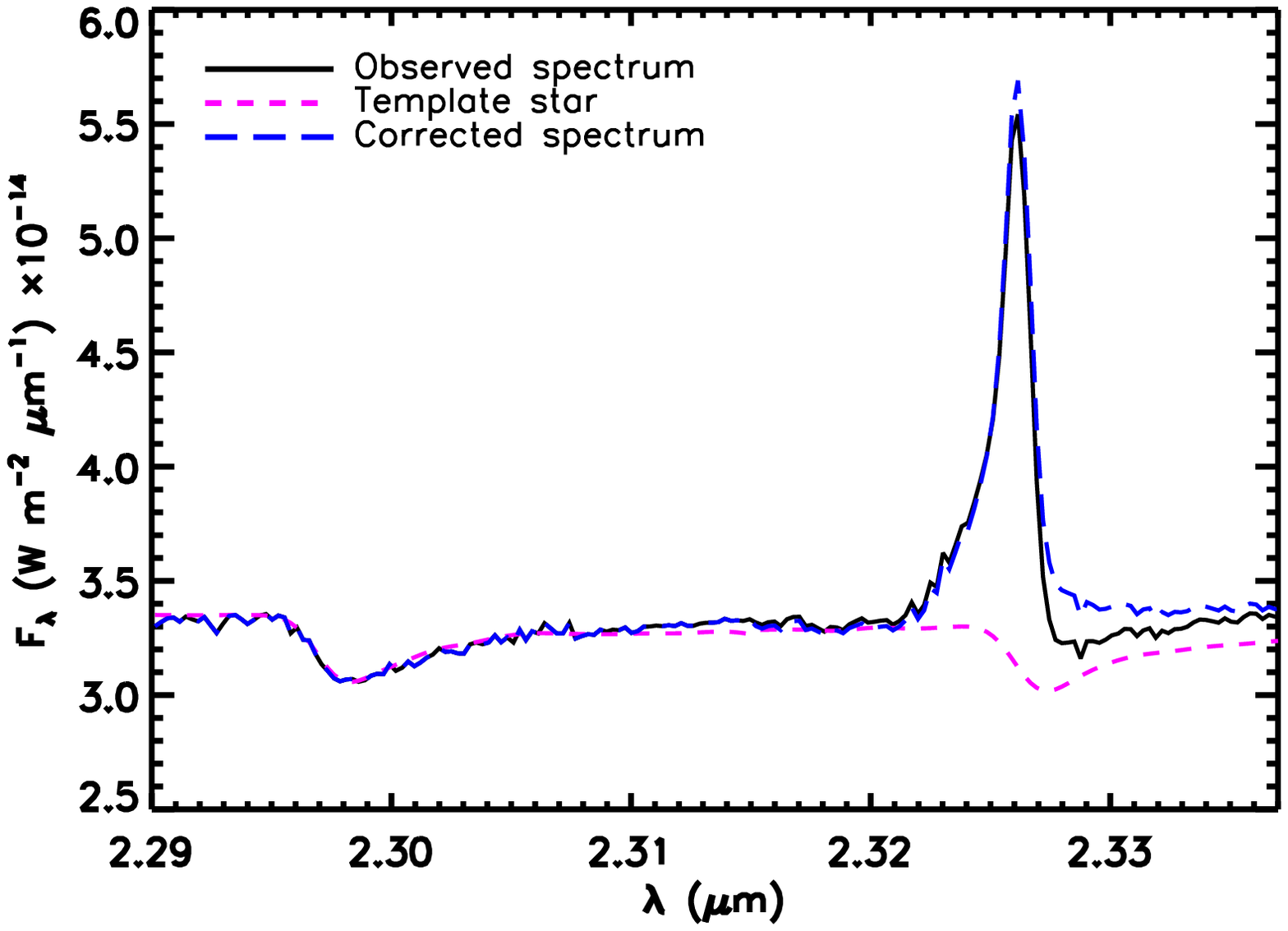}}   
     \hspace{0.5cm}
     \resizebox{0.4\textwidth}{!}{\includegraphics[clip]{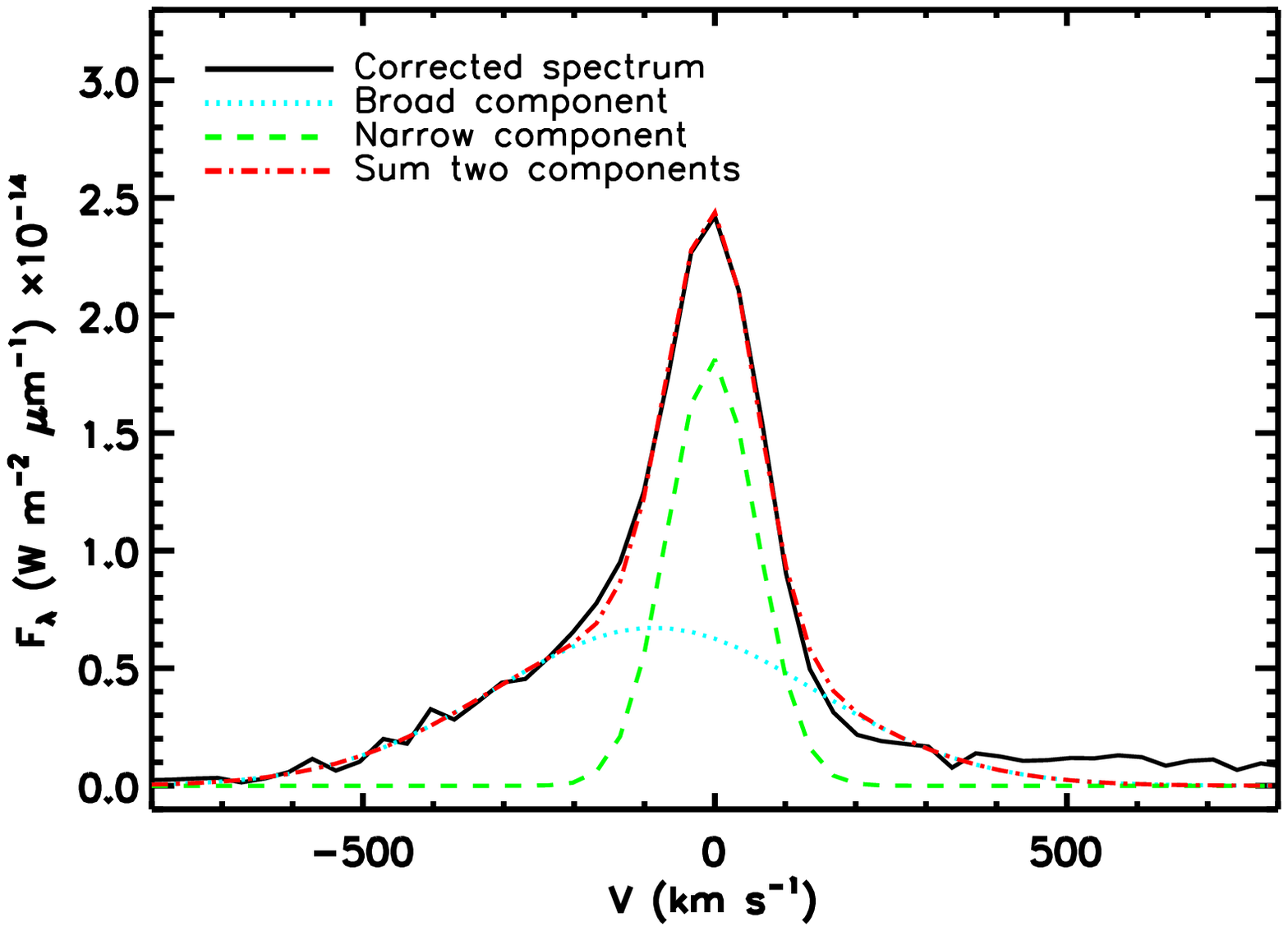}}
    \caption{\textit{Left panel}
             Nuclear spectrum around [Ca{\sc viii}] to show how the stellar
             continuum was substracted.
             \textit{Right panel}  
             [Ca{\sc viii}] profile in velocity space, fitted by two gaussians}
    \label{spectrumcaviii}
    \end{center}
\end{figure*}

Table~\ref{galaxies} lists 17 galaxies for which there are in the
literature both measurements of the H$_2$ 1-0\,S(1) line flux and estimates
of the total gas mass from mm CO\,1-0 or CO\,2-1 luminosities.
The H$_2$ 1-0\,S(1) luminosities are primarily from Goldader et
al. (\cite{gold97}), since these are taken in large (3\arcsec\ or more)
apertures and therefore are more likely to include all, or nearly all,
the H$_2$ 1-0\,S(1) flux.
In addition, these galaxies are actively star forming -- as evidenced
by their classification as luminous or ultraluminous galaxies.
As the table shows, nearly all of these have similar ratios
of the 1-0\,S(1) luminosity with the infrared luminosity $L_{\rm IR}$,
and also with  the molecular mass estimated from the CO luminosity
$L_{\rm CO}$.
The former relation has previously been noted by Goldader et
al. (\cite{gold97}), who suggested it could be explained in terms of
supernova remnants if the H$_2$ originates in gas shocked by the
expanding shells.
On the other hand, Davies et al. (\cite{dav03}) argued that in general the
H$_2$ 1-0\,S(1) line in such galaxies is excited by fluorescence by
young stars of gas dense enough that the lower vibrational levels are
thermalised.
In either case, if the H$_2$ emission is primarily associated with star
formation, it would not be unreasonable to find a correlation with
$L_{\rm IR}$.

Similarly, a correlation between $L_{\rm IR}$ and $L_{\rm CO}$ has
been known for a long time (see Young \& Scoville \cite{you91} for a review).
Recent work (Gao \& Solomon \cite{gao04}) shows that while 
$L_{\rm IR}$ and $L_{\rm CO}$ are correlated, the logarithmic relation
is linear only over limited luminosity ranges and there is
a clear trend of $L_{\rm IR}$/$L_{\rm CO}$ (i.e. star formation
efficiency) with luminosity.
They also showed that if one uses instead a tracer of dense gas such as
HCN, the equivalent ratio $L_{\rm IR}$/$L_{\rm HCN}$ is constant;
although for our application one then has the difficulty of estimating
the gas mass from the $L_{\rm HCN}$.

In both relations, NGC\,6240 is a major exception, having by far the
largest H$_2$ 1-0\,S(1) luminosity.
The reason is most probably that the excitation mechanism is
most probably shocks rather than fluorescence (e.g. Sugai et al. \cite{sug97}).
For this reason, we have excluded NGC\,6240 from the sample.
%In addition to being active star forming galaxies, the sample in
%Table~\ref{galaxies} also include known AGN, making general
%conclusions about the relation between H$_2$ 1-0\,S(1) luminosity and total
%gas mass that much more robust.

Given the correlations above, it would not be
unreasonable also to expect a relation between 1-0\,S(1) luminosity and
H$_2$ mass as traced by the CO luminosity (or even better would be HCN
luminosity).
In fact we find that the logarithmic ratio of these two quantities is 
log$\frac{L_{\mathrm{(1-0)S(1)}}}{M_{\mathrm{gas}}} = -3.6\pm0.32$ which 
corresponds to  
$\sim2.5\times 10^{-4} L_{\sun}$/$M_{\sun}$ with a 1$\sigma$
uncertainty of a factor 2.
This is a quite remarkable result for such a diverse sample -- that
includes galaxies with powerful AGN for which X-ray irradiation of gas
may be an important contributor to the 1-0\,S(1) luminosity.
It gives us confidence that we can use the H$_2$ 1-0\,S(1) line, without
needing to know the details of how it is excited, to make at least an
approximate estimate of the total (cold) molecular gas mass.

We apply this result to the nuclear region of
Circinus because it has had recent vigorous star formation
close to the AGN, and thus matches the sample.
From the total H$_2$ 1-0\,S(1) luminosity of $4.4\times 10^3$ $L_{\sun}$ 
we estimate the total molecular mass to be $1.7\times 10^7$ $M_{\sun}$
-- fully consistent with the dynamical mass estimated in
Section~\ref{kinematics}. 
The gas mass surface density is then 
$8\times 10^4$ $M_{\sun}$ pc$^{-2}$, yielding a column density of 
$n_{\mathrm{H}}$ = $5.2\times 10^{24}$ cm$^{-2}$, which is very similar to that
implied by X-ray observations (e.g. Matt et al. \cite{matt99}). 
Since $n_{\mathrm{H}}$ $\approx$ $1.5\times 10^{21}$ $\tau_{\mathrm{V}}$, 
the mean gas density implies an optical depth through the thickness of
the disk of $\tau_{\mathrm{V}}$ = $3000$. 
This is far larger than the most likely values estimated in
Section~\ref{starformation}, and also inconsistent with the maximum
possible optical depth.

This discrepancy suggests that the canonical torus 
in the unified model of AGN is not only forming stars but is a clumpy
medium  -- so that much of the star light is not obscured despite the
presence of huge gas column densities.

\section{Coronal lines} \label{coronal}

%  
%\begin{figure}
%    \begin{center}
%     \resizebox{\hsize}{!}{\includegraphics[clip]{circinus_fig10a.eps}}   
%    \caption{\textit{Upper panel}
%             Nuclear spectrum around [Ca{\sc viii}] to show how the stellar
%             continuum was substracted.
%             \textit{Lower panel}  
%             [Ca{\sc viii}] profile in velocity space, fitted by two gaussians}
%    \label{spectrumcaviii}
%    \end{center}
%\end{figure}    
% 

%  
\begin{figure}
    \begin{center}
     \resizebox{\hsize}{!}{\includegraphics[clip]{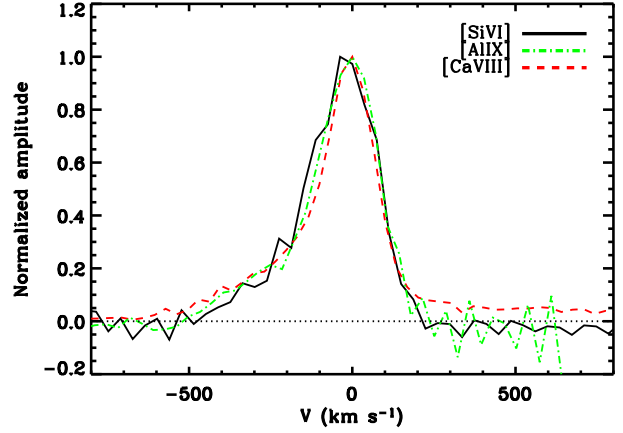}}   
    \caption{Comparison of coronal line profiles in velocity space.}
    \label{clprofiles}
    \end{center}
\end{figure}    
\begin{figure}
    \begin{center}
     \resizebox{\hsize}{!}{\includegraphics[clip]{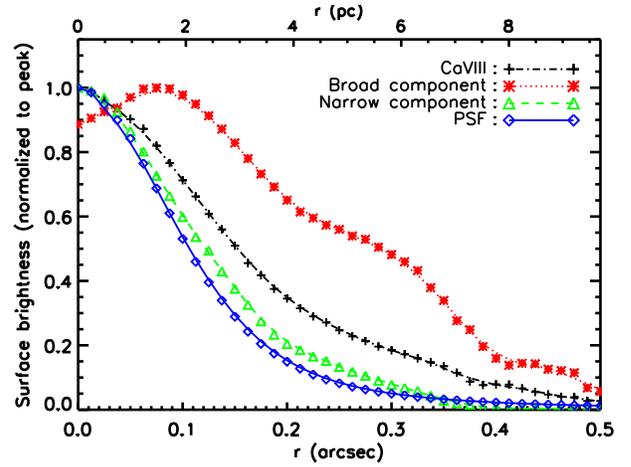}}   
    \caption{Radial profile of the surface brightness of the [Ca{\sc viii}]
            broad component (asterisks connected by a dotted line), 
            and [Ca{\sc viii}] narrow component 
            (open triangles connected by a dashed line), 
            compared to the [Ca{\sc viii}] profile 
            (crosses connected by a dotted-dashed line) and the PSF profile
            (open diamonds connected by a solid line) }
    \label{profiles3}
    \end{center}
\end{figure}    

\begin{table*} 
        \begin{center}
        \begin{tabular}{l c c c c c c}
        \hline 
        \hline \noalign{\smallskip}
        Line & IP & Component & Flux & $F_{\mathrm{broad}}$/$F_{\mathrm{narrow}}$$^a$ &
        Velocity$^b$ & FWHM Line Wdith \\   
           & (eV) & & (W m$^{-2}$$\times 10^{-18}$) & & (km s$^{-1}$) & (km s$^{-1}$) \\ 
        (1) & (2) & (3) & (4) & (5) & (6) & (7)\\
        \hline \noalign{\smallskip} 
        [Si{\sc vi}] & 167 & Broad & 26.8 & &-103 & 300  \\
        $\mathrm{[Si}${\sc vi}] & 167 & Narrow & 24.7 & 1.1 &$\sim$0 & 175  \\
        $\mathrm{[Al}${\sc ix}] & 285 & Broad & 2.6 &  &-228 & 300  \\
        $\mathrm{[Al}${\sc ix}] & 285 & Narrow & 8.2 & 0.3 &$\sim$0 & 200  \\
        $\mathrm{[Ca}${\sc viii}] & 127 & Broad & 28.0 &  &-87 & 540  \\
        $\mathrm{[Ca}${\sc viii}] & 127 & Narrow & 20.0 & 1.4 & $\sim$0 & 150  \\

        \hline
        \hline
        \end{tabular}
        \end{center}

$^a$ The ratio for each line is written in the row of the narrow component.   \\
$^b$ Velocity shift from the systemic velocity.\\

       \caption{Properties of the broad and narrow components of the detected coronal lines.}
\label{clr}
\end{table*}

Because of the high ionization potential (IP$>$100 eV) associated with 
Coronal Line Region (CLR) emission lines, highly energetic processes are required. 
The lines can be excited either by a hard UV to soft X-ray continuum, 
very hot collisionally ionized plasma, or a combination of both.
If the excitation occurs via collisions, the gas temperature should be about 
$T = 10^6$ K. In the case of photoionization via the hard AGN continuum, 
temperatures of only a few $10^3-10^4$ K are needed.

In our observations of Circinus the [Si{\sc vi}], and [Ca{\sc viii}]
lines are extremely strong and [Al{\sc ix}] is also detected, as
apparent in the spectrum in Figure~\ref{spectrumcircinus}. 
Due to the low signal to noise ratio of the $K$-band spectrum after
2.42$\mu$m, we exclude the [Si{\sc vii}] line at 2.48 $\mu$m 
studied by Prieto et al. (\cite{prieto05}).      
From this spectrum and also that in Figure~\ref{spectrumcaviii}, it
can be seen that the [Ca{\sc viii}] 
line sits on top of the stellar $^{12}$\,CO(3-1) absorption bandhead. 
A template star (HD\,179323, shifted to the redshift of Circinus) 
was used to correct the stellar features in the spectrum 
and hence reconstruct the full [Ca{\sc viii}] line profile. 
As the deep CO bandheads are produced in the atmospheres of red (super)giant 
stars which dominate the emission around $2.3\,\mu$m, HD\,179323 (spectral type K0Iab) 
was choosen for this purpose. 
The correction was achieved by convolving the template with a Gaussian
broadening function,
and varying its parameters to minimise $\chi^2$, which was measured
across the $^{12}$CO\,(2-0) bandhead.
The $^{12}$CO\,(3-1) bandhead of the template was then convolved 
with the broadening function and subtracted from the galaxy's spectrum.

Figure~\ref{clprofiles} shows that in all three coronal line profiles
we observe asymmetric and broadened lines, indicating the presence of
two or more components within the coronal line region. 
In order to quantify the profile, we have fitted a superposition 
of two Gaussians to each line, as is summarised in Table~\ref{clr}.
As can be seen in Figure~\ref{spectrumcaviii} for the case of 
[Ca{\sc viii}], there is
a strong narrow component and a weaker blueshifted broad component.
These also have different spatial extents.
Figure~\ref{profiles3} shows azimuthally averaged radial profiles of 
the broad and narrow components of this line, as well as the PSF.
It is clear that the broad component is resolved and spatially more extended.
The flux peaks at about 1.5 pc from the nucleus. 
On the other hand, the narrow component is unresolved and most of its
flux comes from a region of less than 1 pc across and coincident
with the non-stellar $K$-band continuum peak (rather than the peak in
Br$\gamma$, H$_2$ 1-0\,S(1), and stellar continuum). 
%The data in Table~\ref{clr} and Figures~\ref{clprofiles}
%and~\ref{profiles3} suggest that the broad blue wing of the coronal
%line is spatially extended.
%On the other hand, the narrow component, which is at systemic velocity 
%with respect to Br$\gamma$ and H$_2$ 1-0\,S(1), is compact. 
These characteristics suggest that they originate in different regions 
around the AGN, and perhaps are even excited by different mechanisms. 

In the case of the narrow component, the fact that it is compact and
centered on the nucleus suggests that it originates physically close
to the AGN.
Since the lines are narrow and at systemic velocity, these clouds of
ionized material are not out-flowing which is also an indication that
they are excited by photoionization 
(Oliva at al. \cite{oliva94,oliva99}).

In the case of the broad component, the fact that it is blue shifted
indicates that part of the gas must arise in outflows around the AGN,
as has been suggested by Rodr\'iguez-Ardilla et al. (\cite{rod04}).  
Previous work on the coronal line emission in Circinus (Oliva et al.
\cite{oliva94}) had ascribed the excitation to
photoionisation. 
This was in part due to the narrow width of the most prominent component
(typically 175\,km\,s$^{-1}$, from Table~\ref{clr}).
However, the considerable broadening associated with the blue wing 
(FWHM$>$300 km s$^{-1}$) opens the possibility that some
fraction of it might in fact be excited by fast shocks.
The picture of a shock-excited blue component is supported by the results 
of Prieto et al. (\cite{prieto05}),  
who suggested that in addition to photoionization, 
shocks must contribute to the coronal emission.
They proposed shock excitation as an additional energy source to explain 
the extended coronal line emission, as traced by [Si{\sc vii}], 
in a sample of 4 nearby Seyfert 2 galaxies.
Although it could be a single broad component, we consider a more
likely scenario to involve many narrow components arising from
different clouds moving at different
velocities whose emission combines to produce the observed morphology
and profile.
A mechanism by which broad coronal lines are created in out-flowing
cloudlets which have been eroded from the main clouds is an
intepretation which has been proposed for the detailed morphological
and kinematic data available for NGC\,1068 (Cecil et
al. \cite{cecil02}).

It is interesting that we observe a correlation between the IP
and the blueshift, as well as the $F_{\mathrm{broad}}$/$F_{\mathrm{narrow}}$, 
of the forbidden lines.
As the ionization potential increases, the blueshift also
increases and the $F_{\mathrm{broad}}$/$F_{\mathrm{narrow}}$ decreases.
This can be understood if the acceleration of cloudlets away from the
AGN is driven impulsively (which one might expect given the enormous
variability of hard X-ray luminosity in many AGN) so that the
cloudlets with the highest 
velocities will also be closest to the AGN -- and as they travel
further out, they will be decelerated by drag against the interstellar
medium (probably small in the ionisation cone) and by the increasing
gravitational pull.
The gravitational pull is significant since it depends not only on the
black hole mass, as 
assumed by Maiolino et al. (\cite{maio00}), but also on the gas
and stars in the nuclear region which we have shown provide an order
of magnitude more mass than the black hole alone on scales of 10\,pc.
Additionally, those lines which required the hardest ionising continuum
will arise preferentially closer to the AGN.
Thus although the strength of the narrow component which originates
close to the AGN will 
depend on IP since there are fewer photons energetic enough to ionise
species with higher IP;
it will be even harder to create these lines in outflowing
cloudlets, since they are already further from the AGN:
they will arise only in the fastest clouds closest to the AGN.
Lines with less extreme IP will be generated also in slower clouds
further from the AGN and hence the mean blueshifted velocity will be
less.
Although we do not find a clear correlation between the IP and the
width of the fit to the broad component,
that of the [Ca{\sc viii}] line exhibits the highest dispersion for the
lowest IP, providing further support for this hypothesis that lines
with the lower IP can arise in regions which are further from the AGN
and hence moving outward slower.

\section{Conclusions} \label{concl}

We have presented near-infrared adaptive optics integral field
spectroscopy of the nuclear region of the Circinus galaxy with an
angular resolution of 0.2\arcsec\ and spectral resolution $R \sim 4200$,
which we use to probe the gas and stellar morphologies and kinematics on
scales of a few parsecs.

In the central 0.8\arcsec, the non-stellar continuum dominates the
$K$-band, contributing 85\% of the total flux density.
Offset by $\sim$0.15\arcsec\ from its peak are the H$_2$ 1-0\,S(1) and
Br$\gamma$ lines and stellar continuum.
The similarity of their morphologies and kinematics lead us to
conclude that they all originate predominantly with recent vigorous
star formation.
Analysis of these diagnostics together with the radio continuum
suggest that within 8\,pc of the AGN there is a starburst which is
exponentially decaying on a timescale of 100\,Myr and began only
80\,Myr ago, and currently has a bolometric luminosity of $\sim$1.4\%
that of the entire galaxy.

We have found that it is possible to make a reasonable estimate of the
total gas mass from the H$_2$ 1-0\,S(1) line luminosity, finding that it is
an order of magnitude more than the young stars, and also consistent
with the column densities to the AGN implied by X-ray observations.
Since the scales on which the gas and stars exist are similar, we
suggest that the torus is forming stars;
and because of the limits on extinction to the stars, that the torus
must be a clumpy medium rather than uniform.

The coronal lines comprise two components: a prominent narrow 
(FWHM $\sim 175$\,km\,s$^{-1}$) part which is at systemic
velocity, spatially unresolved (at our 4\,pc FWHM resolution), and
centered on the non-stellar continuum;
and a broad (FWHM $>300$\,km\,s$^{-1}$) blue shifted part which
is spatially extended.
We argue that the narrow part arises in clouds physically close to
the AGN; and that the blue wing originates, as appears to be the case in some
other Seyfert galaxies, from cloudlets that have been eroded form the
main clouds and are accelerated outward.

%-------------------------------------------------------------------------------

\begin{acknowledgements}

The authors are grateful to Almudena Prieto for kindly providing the
$K$-band NACO image used for estimation of our spatial resolution.
The authors are most gratefull to the entire SINFONI team from ESO, 
MPE, and NOVA for their support in the SINFONI comissioning.

\end{acknowledgements}

%------------------------------------------------------------------------------------

\end{document}